\documentclass[a4paper,twoside]{article}

\usepackage{epsfig}
\usepackage{subcaption}
\usepackage{calc}
\usepackage{amssymb}
\usepackage{amstext}
\usepackage{amsmath}
\usepackage{amsthm}
\usepackage{multicol}
\usepackage{pslatex}
\usepackage{apalike}
\usepackage[bottom]{footmisc}
\usepackage{enumitem}
\usepackage[ruled,vlined,linesnumbered]{algorithm2e}
\usepackage{array}
\usepackage{wasysym}

\newtheorem{definition}{Definition}

\usepackage{SCITEPRESS}     % Please add other packages that you may need BEFORE the SCITEPRESS.sty package.

\begin{document}

\title{Local Visibility Roadmaps in Continous Multi-agent Path Finding}

\author{\authorname{Kristýna Janovská\sup{1}\orcidAuthor{0000-0001-6350-6357} and Pavel Surynek\sup{1}\orcidAuthor{0000-0001-7200-0542}}
\affiliation{\sup{1}Czech Technical University in Prague, Faculty of Information Technology, Th\'{a}kurova 9, 160 00 Praha 6, Czechia}
\email{\{janovkri, pavel.surynek\}@fit.cvut.cz}
}

\keywords{VR-CBS, MAPF, Visibility Roadmaps, Multi-agent Pathfinding, B-Spline Curves, Conflict-based Search}

\abstract{We address the problem of smooth continuous multi-agent path finding (SC-MAPF), where agents move in a continuous environment along smooth curves in continuous time. The paths for individual agents are searched in local visibility roadmaps that consider obstacles and other agents. A new algorithm is proposed in this work, which combines the previous algorithm CE-CBS for high-level search with any-angle path finding for local single-agent planning, where agents are to avoid obstacles, both static and other agents.  The new algorithm is tested and compared to CE-CBS, the algorithm building individual paths using RRT$^*$, and SMT-CCBS algorithm, an algorithm for MAPF$_R$. Experimental results show various advantages of the new algorithm over CE-CBS and SMT-CCBS in comparable settings.
}

\onecolumn \maketitle \normalsize \setcounter{footnote}{0} \vfill

\section{\uppercase{Introduction}}
\label{sec:introduction}

{\em Multi-agent path finding} (MAPF) is a problem of navigating agents in a shared environment from their initial positions to given individual goal positions while agents must not collide with each other.

Recent development in the field of multi-agent path finding has seen a shift from purely discrete agent-based modeling \cite{bonabeau}, where the environment is represented as an undirected graph $G=(V,E)$ with agents represented via occurrences in $V$, to works containing various degrees of realism -- from continuous time and space \cite{WalkerSF18} to implementing kinematic constraints for agent movement \cite{HonigK00XAK17} such as the necessity of smooth movement, which can be an advantage for agents in terms of speed changes necessary during sharp turns \cite{curvesRRT}.

Some recent work such as \cite{ccbs} and \cite{Yan_2024} has focused on MAPF with a continuous time but discrete environment. A collision between a pair of agents is resolved in the time domain by making one of the agents participating in the collision to wait. In contrast to this, we address not only continuous time but also a continuous environment so that agents have a chance to change their trajectory to avoid the collision. We present a new model combining the previous {\em continuous environment conflict-based search} \cite{cecbs} and visibility roadmaps, in which we construct an individual visibility roadmap for each agent according to its constraints (derived from previous collisions) and static obstacles. The paths found on these roadmaps are then smoothed by B-spline curves \cite{nurbs} to obtain trajectories that can be followed by real robots more easily. 

The problem of MAPF often serves as a basic abstraction for the motion planning of multiple robots such as UAV swarms \cite{drones} or warehouse robots \cite{PrabhuC21}. Hence, the smooth trajectories generated by the continuous MAPF approach contribute significantly to the faster execution of robot movements.

\subsection{Contribution}
In this work, the VR-CBS algorithm is proposed. The VR-CBS algorithm resolves collisions between agents in the space domain similarly as it is done in CE-CBS. In contrast to CE-CBS, VR-CBS uses per-agent visibility roadmaps derived from both static obstacles and other agents. VR-CBS is a two level CBS-based \cite{SharonSFS12} algorithm, where the low level of trajectory search is performed in the visibility roadmap and the high level ensures search across all alternatives of collision resolutions.

\section{\uppercase{Related work}}
In \cite{cecbs}, a novel algorithm CE-CBS for MAPF in continuous environment was proposed. It combined a high-level CBS-based approach with random sampling provided by low-level RRT* path finding. Elimination of collisions between agents is done in the space domain by making one of the agents participating on the collision to make an evasive maneuver. Paths obtained from RRT* are smoothed with B-spline curves in CE-CBS to achieve more realistic trajectories. This approach showed how multiple agents can plan smooth collision-free paths in continuous time and space.

The idea of {\em Sparse Visibility Graph} search is studied in \cite{ensvg}, where an optimal approach for any-angle path finding is proposed. The algorithm constructs a visibility graph \cite{VG} using line-of-sight checks and taut region checks. The visibility graph is further optimised to make path finding more efficient. Edges of the visibility graph are pruned in a way that optimality is preserved, while running time is significantly improved thanks to pruning.

In \cite{YakovlevAS24}, optimal any-angle MAPF approach is proposed. The work is based on {\em Continuous Conflict-Based Search} (CCBS), with the lower level of this algorithm being an any-angle variant of Safe Interval Path Planning, which provides optimal solutions. This method is then optimised using disjoint splitting and multi-constraints. In addition, a bounded-suboptimal variant of this algorithm is proposed, which optimises the runtime while keeping solutions suboptimal.

In \cite{liang2024}, the authors approached the problem of multi-robot motion, proposing a model using multi-model diffusion. Data-driven smooth paths were generated on the basis of conflict elimination constraints while combining several diffusion models to generate effective paths. Diffusion models for a single robot can be effectively combined with classical MAPF algorithms, resulting in multi-robot trajectories.

The authors of \cite{okumura2022} have also proposed a generative model learning approach, developing cooperative timed roadmaps (CTRM). Agents in these roadmaps focus on important locations in the environment, while avoiding collisions with other agents. A generative model is learnt from numerous instances and solutions and is able to generate new problem instances.

In \cite{Kottinger_2022}, the K-CBS algorithm is introduced as a solution to multi-robot motion planning, where agents not only plan non-conflicting trajectories but are also subject to kinodynamic constraints. The proposed algorithm works in a continuous domain and uses an initially decentralised approach, in which initial paths are planned and constraints imposed, and transforms into a semi-centralised approach as it progresses and merges some robots into a meta-robot. The algorithm is probabilistically complete.

\section{\uppercase{Background}}

The problem of multi-agent path finding with both continuous time and space while using smoothing mechanisms corresponds to that defined in \cite{cecbs}:

\begin{definition}[Smooth Continuous MAPF - SC-MAPF]
    Let $\mathcal{M}$ be a metric space, and let $R = \{r_1, r_2, ..., r_v\}$ be a set of agents.

  SC-MAPF is the tuple $\mathcal{E}=[\mathcal{G}=(V,E),\mathcal{M}, S, G,$ $coord, \mathcal{A}]$, where $\mathcal{G}$ represents a graph, $ \mathcal{M}$ is a metric space, $S: R \rightarrow V$ the injective start function and $G: R \rightarrow V$ the injective goal function with $coord$ mapping every vertex in $\mathcal{G}$ to coordinates in $\mathcal{M}$ and $\mathcal{A}$ being a finite set of possible move actions. Every edge $e \in E$ represents a set of smooth polynomial curves $C = \{C_1, ..., C_n\}$ that connect two vertices in $\mathcal{G}$. The set of curves belonging to an edge $e \in E$ is denoted as $C_e$. We assume that these curves can be parametrized by a single real parameter representing time that can be translated by the parameterization to a position on the curve.
 
   Every action $a \in \mathcal{A}$ in MAPF$_R$ is defined by a duration $a_D$ and a motion function $a_\varphi: [0, a_D]: \rightarrow \mathcal{M}$ that maps time to metric space. This motion function corresponds to a smooth curve $c_i \in C_e$ and is derived from the parameterization of $c_i$. $a_\varphi (t)$ are the coordinates of an agent in the metric space $\mathcal{M}$ at a time $t$ while executing an action $a$. 

 A problem of {\em smooth continuous multi-agent path finding} (SC-MAPF) is the task of finding a set of smooth curves -- a sequence of non-conflicting actions, each choosing a curve $c_i$ from a set of curves $C_e$ belonging to a respective edge $e$. 

\end{definition}

Next we define the solution to this problem, which has several constraints regarding the agent speed, trajectory and collisions between two differnt agents.

\begin{definition}[SC-MAPF Solution]
The solution to the problem of SC-MAPF is constructed while satisfying the following constraints:

 \begin{enumerate}[label=(\roman*)]
\item Every agent $r_i$ moves at a constant speed.
\item The trajectory of an agent is a smooth parametric polynomial curve $c$, which satisfies the given kinematic constraints.
\item A sequence of actions $A$ can be valid only if no two agents collide, meaning no two agents' bodies overlap at any time.
\end{enumerate}
\end{definition}

In this work, the agent body is a circle defined by the coordinates of its center and its radius $r$. Although all agents share the value of radius $r$ in experiments presented here, varied values of $r$ can be used for each agent.

\begin{definition}[Agent body]
The agent body is defined as a shape in the metric space $\mathcal{M}$, usually a convex polygon or a circle.
\end{definition}

Lastly, a conflict in smooth continuous MAPF is defined.

\begin{definition}[SC-MAPF Conflict]
A conflict is defined as a 6-tuple $\mathcal{C} = [r_i, a(r_i), r_j, a(r_j), t(r_i), t(r_j)]$, where $r_i$ and $r_j$ are agents involved in the conflict, $a(r)$ is the action agent $r$ performs for $r \in \{r_i,r_j\}$, and $t(r)$ is the time when an agent $r$ starts carrying out the action $a(r)$ leading to the conflict for $r \in \{r_i,r_j\}$. A conflict means that the bodies of agents $r_i$ and $r_j$ overlap when agents perform actions $a(r_i)$ and $a(r_j)$ starting at time $t(r_i)$ and $t(r_j)$, respectively. Solving a conflict means forbidding the agents to move on certain curves which would traverse through the conflicting position at a conflicting time.
\end{definition}

\section{\uppercase{Proposed model}}
Our new algorithm, called {\em Visibility Roadmap Conflict-Based Search} (VR-CBS), consists of two base parts ---- the high level is inspired by that of the previous algorithm CE-CBS \cite{cecbs}, while the lower level single agent path finding is done by first constructing a visibility roadmap \cite{VG} of the environment considering the agent's constraints. The search in the visibility roadmap is done via A* \cite{hart1968a}, and the resulting path is smoothed using B-spline curves \cite{nurbs}. The algorithm fits our definition of SC-MAPF where the set of parametric curves associated with the edges of the input graph $\mathcal{G}=(V,E)$ is represented by B-spline curves connecting the endpoints of the edges in the metric space $\mathcal{M}$. We also assume circular agents of identical size.

\subsection{\uppercase{High-level VR-CBS}} In the higher level of VR-CBS, shown in Algorithm \ref{alg:vrcbs}, a CBS-like conflict search routine is performed. In line 1, an initial set of paths is found in roadmaps constructed for each agent separately with the visibility roadmap search function. This function is further described in the next section.

The constructed roadmaps are unique for each agent $r$, as they are constructed with line-of-sight check from their starting positions and also work with their respective constraints, according to which these roadmaps are updated once a new VR-CBS node is formed. A constraint is
defined as a 4-tuple $c = (r, p, \mathit{unsafe}_{\mathit{from}}, \mathit{unsafe}_{to})$, where $r$ specifies an agent to which that constraint applies, $P$ is a coordinate of a conflict and $\mathit{unsafe}_{\mathit{from}}, \mathit{unsafe}_{to}$ denotes the unsafe time interval. If an agent crossed $p$ in a time within the unsafe interval, it would mean breaking that constraint.

Conflicts from which constraints are constructed can have two causes:
\begin{enumerate}
    \item If two agents' trajectories cross, the intersection point is a possible point of conflict. Unsafe intervals are constructed for both agents, which represent the time interval in which each of these agents overlaps with the intersection point. If the unsafe intervals intersect, the agents are in a conflict.
    \item Two trajectories do not necessarily have to cross for the agents to collide. As they are rigid bodies, it can suffice for the trajectories to get close to each other without intersecting. If two trajectories have at two points distance between them lower than twice the agent radius, these points are possible points of conflict. Unsafe intervals and their comparison is done as in the first case. In this case, however, since the trajectories do not intersect, each agent will have its own conflicting position. 
\end{enumerate}

In each iteration of VR-CBS, a solution node of the CBS solution tree is expanded \cite{cbs}. One solution node contains a set of paths which can be conflicting and a set of constraints for agents to which these paths belong. A solution to expand is selected prioritising those nodes with the shortest sum of costs (SOC) of all solutions, as seen in line 4. In line 6, a first conflict in a solution is found with a time check which decides if two agents overlap at a time and selects the first moment when they do so. For this tuple of agents, two new sets of solutions are found, each updated with one constraint belonging to one of the conflicting agents (lines 9-13). These solutions become two new solution nodes. These solution nodes can be expanded next, depending on their SOC. The search ends once a conflict-free path is found or when the instance is deemed unsolvable, which is made possible by the construction of visibility roadmaps with permanent agent obstacles. This is discussed further in the next section. However, both because of these permanent agent obstacles and the smoothing mechanism, which may not always be able to return a valid solution, some valid solutions may be omitted.

\begin{algorithm}[h!]
\SetKwInOut{Input}{Input}
\SetKwInOut{Output}{Output}
\Input{agents $R=\{r_1, ..., r_n\}$, obstacles O}
\Output{non-conflicting paths $\Pi=\{\pi_{r_1}, ..., \pi_{r_n}\}$}

$init$\_$solution \leftarrow$ visibility$\_$roadmap$\_$search(R, O)\;
open $\leftarrow$ $\{(init$\_$solution, \emptyset)\}$\;

\While{open $\neq \emptyset$}{
    $(best$\_$solution, constr)$ $\leftarrow$ min$\_$soc$\_$sol(open)\;
    open.remove($(best$\_$solution, constr)$)\;
    conflict $\leftarrow$ first$\_$conflict($(best$\_$solution, constr)$)\;
    \If {conflict = $\emptyset$}{
        \Return $(best$\_$solution, constr)$\;
        }
    \For{${r_i}$ in conflict.agents}{
        $[u_{from}, u_{to}]$ $\leftarrow$ get$\_$unsafe$\_$interval($r_i$, conflict)\;
        new$\_$constr $\leftarrow$ constr $\cup \{(r_i, \mathit{position}_{r_i}, u_{from}, u_{to})\}$\;
        
        new$\_$solution $\leftarrow$ update$\_$solution($r_i$, $(best$\_$solution, new\_constr)$)\;
        open $\leftarrow$ open $\cup$ $\{(new\_solution, new\_constr)\}$\;
    }
}

\caption{Higher level of VR-CBS}
\label{alg:vrcbs}
\end{algorithm}

\subsection{Low-level path finding}
This section describes the lower level of VR-CBS, which focuses on path finding for a single agent while respecting the given constraints. This section is structured as follows: First, the construction of a visibility roadmap used for path finding is described, including constraint solving. Then, the path finding with path smoothing is described.

The low-level path finding problem finds a solution for all agents $R = \{r_1, r_2, ..., r_v\}$ or an agent $r_i$ with new constraints, as shown in \ref{alg:vrcbs} in lines 1 and 12. In line 1, an initial solution without any constraints is found for every agent. After a conflict is found, two new solutions are constructed by updating the path of one of the agents present in the conflict. This is done on line 12 by the function $update\_solution$. 

When searching for a solution for a single agent $r_i$, a visibility roadmap is built first. A visibility roadmap is a variation of a visibility graph \cite{VG} in a continuous environment, defined as follows. 

\begin{definition}[Visibility Roadmap]
Let $VR = (N, L)$ be an undirected graph and let $\mathcal{M}$ be a metric space with $N \subset \mathcal{M}$. $VR$ is a visibility roadmap, when the vertex set $N = O \cup S \cup G$, where $O$ is the set of all coordinates of the corners of obstacles, $S$ are the coordinates of the starting vertex, and $G$ are the coordinates of the goal vertex, and $L$ is the set of all links $(n_i,n_j)$.
\end{definition}

Obstacles in this work are defined as convex polygons. A link is defined as a straight line connecting the $i$-th element of $N$ to the $j$-th element, which does not overlap any obstacle.

\begin{figure}[h!]
\includegraphics[scale=0.5]{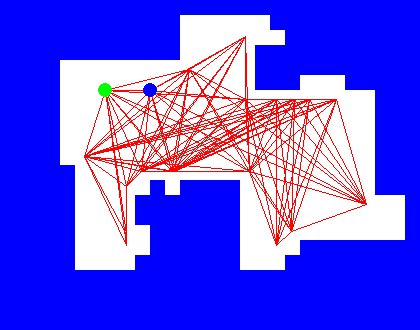}
\caption{Visualization of a visibility roadmap in a continuous map for agent with start position marked as a blue circle and goal position marked as a green circle. Vertices created from obstacle corners have added padding corresponding to agent radius $r$. This map is a converted environment originally introduced in \cite{movingai}.}
\label{fig:visibility_roadmap}
\end{figure}

First, the vertices of this graph are initialised. This part can be seen in Algorithm \ref{alg:vrcbs-init}. All possible vertices are either obstacle corners, agent's start and goal positions, and corners of agent obstacles. Agent obstacles are hexagonal approximations of an agent body at a certain position, which is tied to a constraint passed to low level from the higher level of this algorithm.

A hexagonal shape was chosen based on preliminary experiments as a trade-off between accuracy of the representation leading to possible suboptimality of the solution and number of vertices constructed, which, if high, would have negative effects on the runtime.

\begin{figure}[h!]
\includegraphics[scale=0.24]{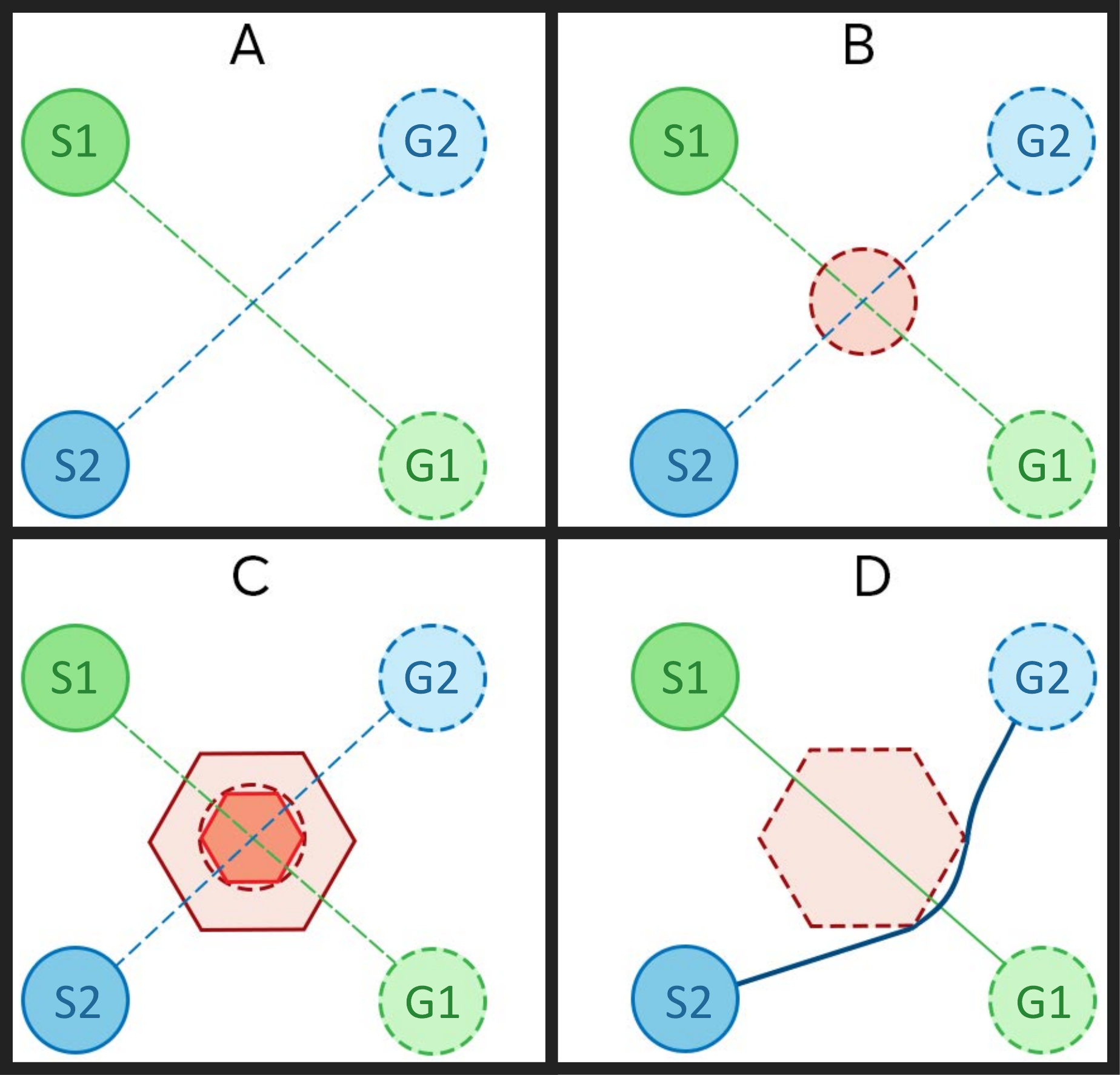}
\vspace{0.5cm}
\caption{Figure visualizes a conflict and its resolution in four steps - (A) conflict, (B) unsafe segment creation, (C) hexagonal obstacle creation in  function $convert\_constraint$ in Algorithm \ref{alg:vrcbs-init} and (D) visualizes final path smoothing and path adjustment.}
\label{fig:4}
\end{figure}

In Figure \ref{fig:4}, part (C), the process of solving these constraints is visualised. In Algorithm \ref{alg:vrcbs-init}, this process is done on line 8 in function $convert\_constraints$, which constructs new vertices in the visibility roadmap based on constraints. In picture (A), it is shown how two agents would collide in the middle of their paths. The' starting positions of the agents are marked as $S1$ and $S2$, the goal positions are $G1$ and $G2$. In (B), an unsafe segment (in red) is created for the blue agent $r_{blue}$, with a red circle in the middle with radius $\rho$ of the agents. The blue agent is given a constraint $c = (r_{blue}, p, \mathit{unsafe}_{\mathit{from}}, \mathit{unsafe}_{\mathit{from}})$. The red circle represents all possible places where $r_{blue}$ would touch or overlap with the intersection point $p$, while being centered on its path, shown in blue. The red region, which intersects the blue path, thus describes all coordinates, where $r_{blue}$ would traverse while in the unsafe interval $(\mathit{unsafe}_{\mathit{from}}, \mathit{unsafe}_{\mathit{from}})$.

Unsafe intervals are constructed in the following manner: First, an intersection of two trajectories, or closest points, is found. Then, for each agent an interval is computed, which represents the time, when any part of the agent body overlaps with the intersection point. This interval is the unsafe interval. If two agents' unsafe intervals overlap, the agents are in a conflict. The whole unsafe interval is then used to make a constraint.

The red circle itself represents the conflict spot with its immediate surroundings of the size of the agent radius $\rho$, shown in red. In picture (C), a static hexagonal obstacle is formed in place of this conflict circle, the length of a segment of size $2r$ to prevent the agent from touching the red circle. In picture (C), the hexagon is the bigger hexagon shown. Vertices of this hexagon are then added to the visibility graph.

The vertices are connected by a line-of-sight check \cite{ensvg}. Starting from the vertex marking agent's starting position, an edge is created with vertices that are visible from the starting vertex, that is, a possible edge does not intersect an obstacle. Vertices that are visible are examined next, until a visibility graph is formed.
\begin{algorithm}[h!]
\SetKwInOut{Input}{Input}
\SetKwInOut{Output}{Output}
\Input{agent $R$, obstacles $O$, constraints $C$, padding $p$}
\Output{a set of vertices $V$}
vertices $\leftarrow$ agent.start\;
\For{o in O}{
    \For{corner in o}{
        corner.add$\_$padding(p)\;
        vertices $\leftarrow$ vertices $\cup$ $\{corner\}$\;
    }
}
\For{c in C}{
    \If{c.agent = agent}{
        points $\leftarrow$ convert$\_$constraint(c.position, p)\;
        vertices $\leftarrow$ vertices $\cup$ $\{points\}$\;
    }
}
vertices $\leftarrow$ agent.goal\;
\Return vertices\;
\caption{Visibility roadmap vertex creation}
\label{alg:vrcbs-init}
\end{algorithm}

\begin{figure}[h!]
\includegraphics[scale=0.38]{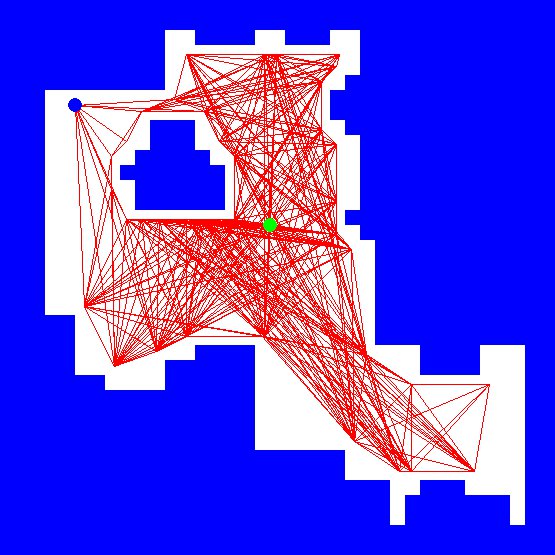}
\caption{Visualization of a visibility roadmap in a continuous map for agent with start position marked as a blue circle and goal position marked as a green circle. This map is a converted environment originally introduced in \cite{movingai}.}
\label{fig:visibility_roadmap_density}
\end{figure}

The density of the visibility roadmap edges depends on the number of convex corners of obstacles. It can be seen in Figure \ref{fig:visibility_roadmap_density}, that the visibility roadmap is more sparse in the part of the environment where the start position (blue circle) is located. As an obstacle is present between the start and goal position (green circle), most obstacle corners are not visible from the start position, and a roadmap must be constructed around it. Once vertices in the larger free space are reached, the roadmap becomes denser as more possibilities arise with more options being visible for every vertices. 

Some vertices, which could be valid, are pruned during the construction of the visibility roadmap, as padding is added into the computation. If a narrow space is present and an agent would not be able to fit in a position without colliding, that vertex is not connected to the visibility roadmap. Vertices can be pruned this way also with the creation of agent obstacles, which can block some of the environment's static obstacles.

\begin{figure}[h!]
\includegraphics[scale=0.38]{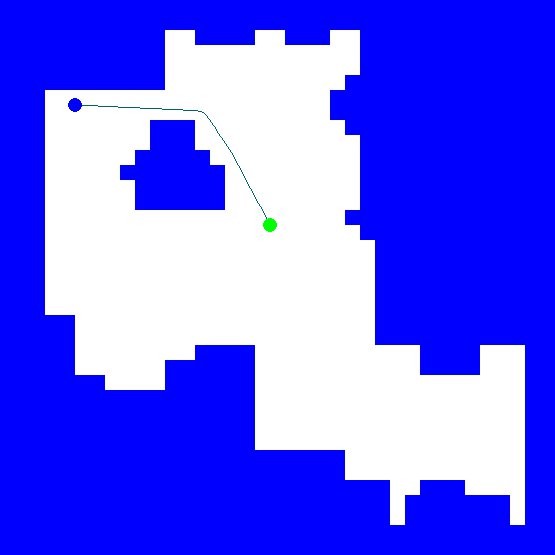}
\caption{A path on map $den201d$ constructed on the visibility graph in Figure \ref{fig:visibility_roadmap_density} and smoothed.}
\label{fig:visibility_roadmap_density_path}
\end{figure}

The hexagonal agent obstacles are permanent and not time-dependent, resulting in conservative conflict solving, which may omit some edge cases. However, including constraints in the visibility graph can help detect when a problem is unsolvable. After adding a constraint, the visibility roadmap for the affected agent is updated. If the result of this update is a visibility graph that does not connect start and goal positions, this solution is marked invalid, and this node of the VR-CBS conflict tree is not further expanded.

\begin{algorithm}[h!]
\SetKwInOut{Input}{Input}
\SetKwInOut{Output}{Output}
\Input{agent $a$, roadmap $VR$, obstacles $O$, constraint $c$, padding $p$}
\Output{path $p$}
constr$\_$vertices $\leftarrow$ convert$\_$constraint(c.position, p)\;

\For{v in $VR$.vertices}{
    \If{collides(c, v)}{
        $VR$.vertices $\leftarrow$ $VR$.vertices $\setminus$ $\{v\}$\;
    }
}
\For{e in $VR$.edges}{
    \If{collides(c, e)}{
        $VR$.edges $\leftarrow$ $VR$.edges $\setminus$ $\{e\}$\;
    }
}
\For{point in constr$\_$vertices}{
    \If{point not in $VR$.vertices}{
        $VR$.vertices $\leftarrow$ $VR$.vertices $\cup$ $\{point\}$\;
    }
}
$VR$.edges $\leftarrow$ $VR$.edges $\cup$ $\{visible\_edges(constr$\_$vertices)\}$\;
$p$ $\leftarrow$ a$\_$star($VR$, $a$)\;
\Return $p$\;
\caption{Update visibility roadmap}
\label{alg:vrcbs-update}
\end{algorithm}

As this graph represents a 2D environment, A* with Euclidean distance as heuristic function is used to find a path. The resulting path is then smoothed with a B-Spline curve. This is shown in picture (D) in Figure \ref{fig:4}. To ensure that the resulting curve does not significantly differ from the original path, all edges are segmented to a set of sub-edges according to a maximal segment length constraint. This allows for precise smooth turns around obstacles. 

The smoothing step can make a path suboptimal. It is, however, included as in a real-life scenario, where an agent is a physical robot, this robot can be sensitive to changes in acceleration, which would occur when the robot was to suddenly change its direction, as is the case of a sharp turn. Hence why differences in closely laid path segments should be limited. \cite{curvesRRT,DTSP,faiglDTSP}

When a new constraint is added, the visibility graph is updated accordingly. This can be seen in Algorithm \ref{alg:vrcbs-update}. First, vertices and edges in the original visibility roadmap, which now collide with the new constraint, are removed from the roadmap. Then, new vertices corresponding to the new agent obstacle are added to the roadmap and connected via a local line-of-sight check. Finally, the A* algorithm is run again to find a path in the new roadmap.

\section{\uppercase{Experiments}}

In this section, we discuss the performance of VR-CBS and compare it to two algorithms - fully continuous CE-CBS and time continuous SMT-CCBS.
We focus on the quality of the solutions produced by VR-CBS and its runtime in the experimental evaluations.

The main metrics used to compare the quality of the solutions in MAPF is the sum of costs (SOC) of all individual paths \cite{SharonSGF13}. This metric is optimised during the path finding in VR-CBS.

Optimal paths are found via A* on individual visibility roadmaps, but since these paths are then smoothed, the resulting paths are suboptimal. The length $\sigma$ of the path of an individual agent is measured as the sum of the lengths of the path curves.

\paragraph{Comparison between VR-CBS and CE-CBS} As both VR-CBS and CE-CBS can solve variants of SC-MAPF - more precisely, both algorithms explicitly use the space domain to resolve collisions between agents - they are candidates for experimental comparison.

\begin{figure}[h]
\includegraphics[scale=0.4]{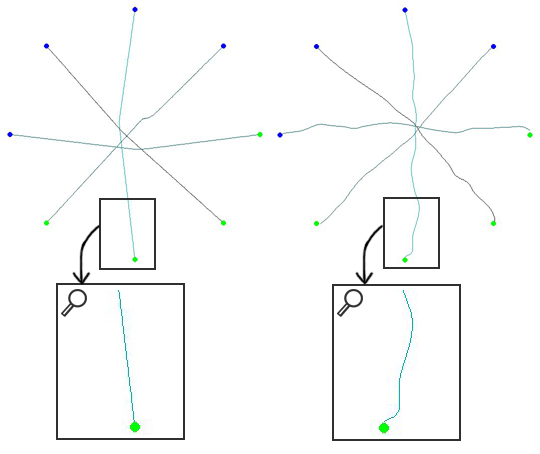}
\caption{A comparison between VR-CBS (left) and CE-CBS(right).}
\label{fig:1}
\end{figure}

\begin{figure}[h]
\includegraphics[scale=0.39]{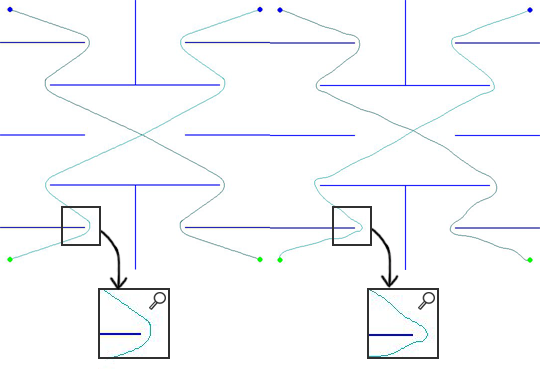}
\caption{A comparison between VR-CBS (left) and CE-CBS(right).}
\label{fig:2}
\end{figure}

Three instances are shown here to demonstrate the difference in how a collision is resolved using an evasive manoeuvre in space between in VR-CBS and CE-CBS. Figure \ref{fig:1} shows an obstacle-free map with four agents. If an optimal path were planned for each agent separately, they would all collide at the same time in the middle of the map. Figure \ref{fig:2} shows a maze map with two agents that could cross in the middle.

This map tests the quality of solutions not only w.r.t to SOC but also w.r.t. curvature (only visual comparison is done).

\begin{figure}[h]
\includegraphics[scale=0.38]{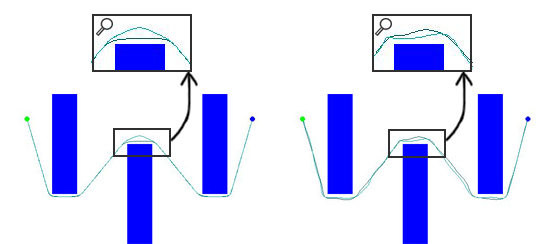}
\caption{A comparison between VR-CBS (left) and CE-CBS(right).}
\label{fig:3}
\end{figure}

The last instance shown in Figure \ref{fig:3} presents an environment with multiple obstacles, where two agents are to switch places. As they meet in the middle part of the map, one agent has to take a detour so that they do not overlap at any time.

\begin{table}[h!]
    \centering
\begin{tabular}{ |c||c|c| } 
\hline
  Instance & VR-CBS SOC & CE-CBS SOC \\  \hline \hline
 Figure \ref{fig:1} - 4 agents & 1992.2 & 2025 \\ \hline
 Figure \ref{fig:2} - maze & 2034.5 & 2062 \\ \hline
 Figure \ref{fig:3} - switch  & 1553.6 & 1632 \\ 
 \hline
\end{tabular}
 \caption{Comparison of SOC of CE-CBS and VR-CBS.}
\label{tab:1}
\end{table}

As can be seen in Table \ref{tab:1}, VR-CBS was able to achieve shorter paths in all instances. As VR-CBS does not use sampling, which in CE-CBS leads to ragged path segments with unnecessary turns, it is able to achieve much straighter and therefore shorter paths while still satisfying all angle constraints.

\paragraph{Comparison of VR-CBS and SMT-CCBS}

VR-CBS was compared on 16 instances with SMT-CBS \cite{SurynekSwarms}, in particular a variant SMT-CCBS \cite{smtcbsr}, which operates with continuous time, that is, the time domain is used for resolving collisions. SMT-CCBS is a reduction-based solver where the reduction to a propositional satisfiability instance (SAT) \cite{handbook_sat} and its solving by a SAT solver is used instead of the CBS framework. 

The instances used for the experiments are based on DAO grid maps \cite{movingai}.

These originally discrete grid maps for the use of VR-CBS were converted to a continuous environment with a greedy algorithm, which merged all 1x1 square obstacles into as big rectangles as possible. This was done to reduce computational time so that collisions are not computed separately on each cell.

The start and goal positions of the agents were randomly generated, with 1 to 10 agents present in each instance. The results can be seen in Table \ref{tab:smt1} and Table \ref{tab:smt2}. During these experiments, both the makespan and sum of costs were evaluated. The {\em makespan}, defined as the length of the longest path of all agents, showed little differences between the two algorithms, with VR-CBS (with the exception of the map $lak108d$ with 10 agents) achieving slightly lower makespan.

Differences in performance are more visible when measuring the sum of costs (SOC), a sum of the path lengths of all agents. In all instances, VR-CBS was able to achieve shorter paths.

The results in Table \ref{tab:smt1} and Table \ref{tab:smt2} show the VR-CBS paths found in cases where 10 agents were present. In all of these cases, it can be observed that due to the construction of visibility roadmaps, agents often try to stay as close to obstacles as possible; mostly if present in a narrow environment. A change of direction appearing on a path that is not caused by the presence of a static obstacle is the result of conflict solving, and in that moment, the agent is actively avoiding an agent obstacle.

\paragraph{Evaluation on narrow corridors}

\begin{figure}[h]
\includegraphics[scale=0.29]{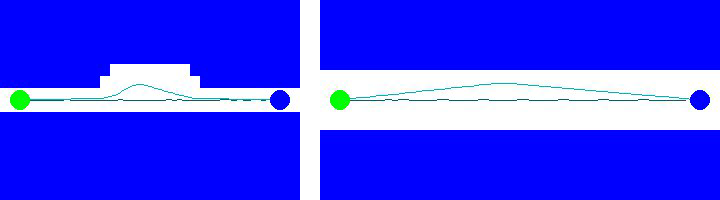}
\caption{VR-CBS solution on two types of a narrow corridor map.}
\label{fig:6}
\end{figure}

The model was evaluated in instances representing narrow corridors. These instances are shown in Figure \ref{fig:6}. In both of these cases, two agents (green and blue) swap their positions. On the left, the corridor itself is as wide as an agent, so a switch would be impossible, but because of the widened part in the middle, one of the agents can take a detour through it and avoid the other agent. On the right, the corridor is wider than the agents, making a swap possible in the corridor itself. Agents therefore demonstrate the ability to navigate even in instances containing these narrow corridors.

\begin{table*}[p] 
\centering
\renewcommand{\arraystretch}{1.2}
\setlength{\tabcolsep}{2pt}
\begin{tabular}{| m{0.14\linewidth} | m{0.34\linewidth} || m{0.14\linewidth} | m{0.34\linewidth}|}

\hline
\multicolumn{2}{|c||}{\textbf{den201d}} & \multicolumn{2}{c|}{\textbf{den202d}} \\[0.3em]
\hline
\includegraphics[width=0.14\textwidth]{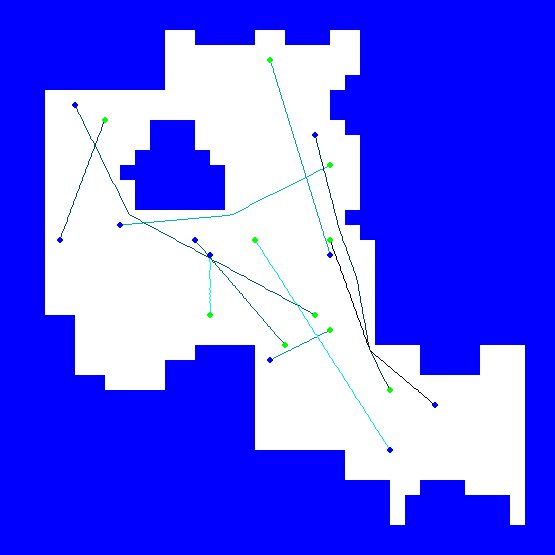} &
\includegraphics[width=0.34\textwidth]{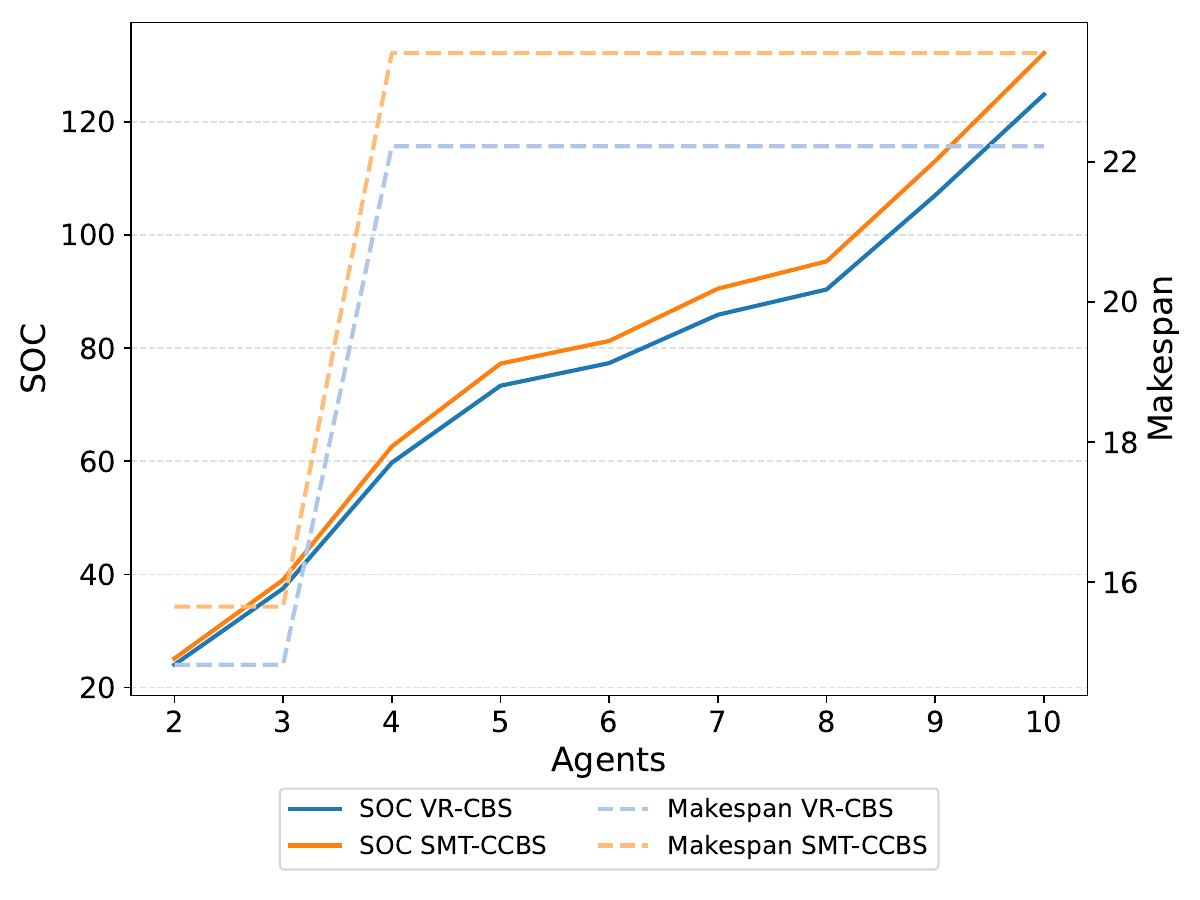} &
\includegraphics[width=0.14\textwidth]{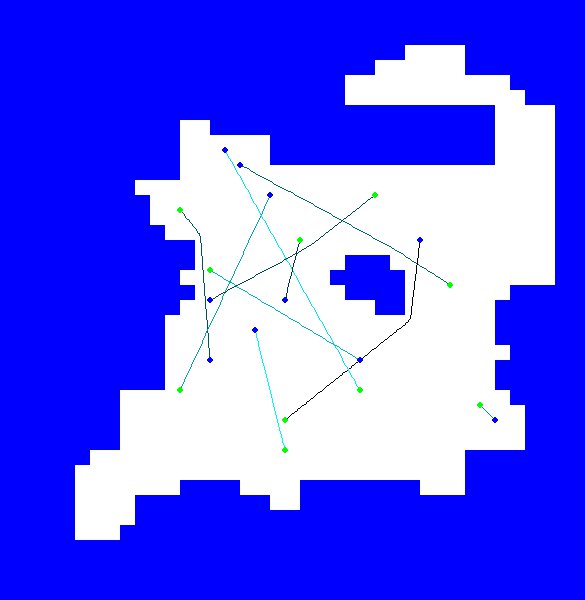} &
\includegraphics[width=0.34\textwidth]{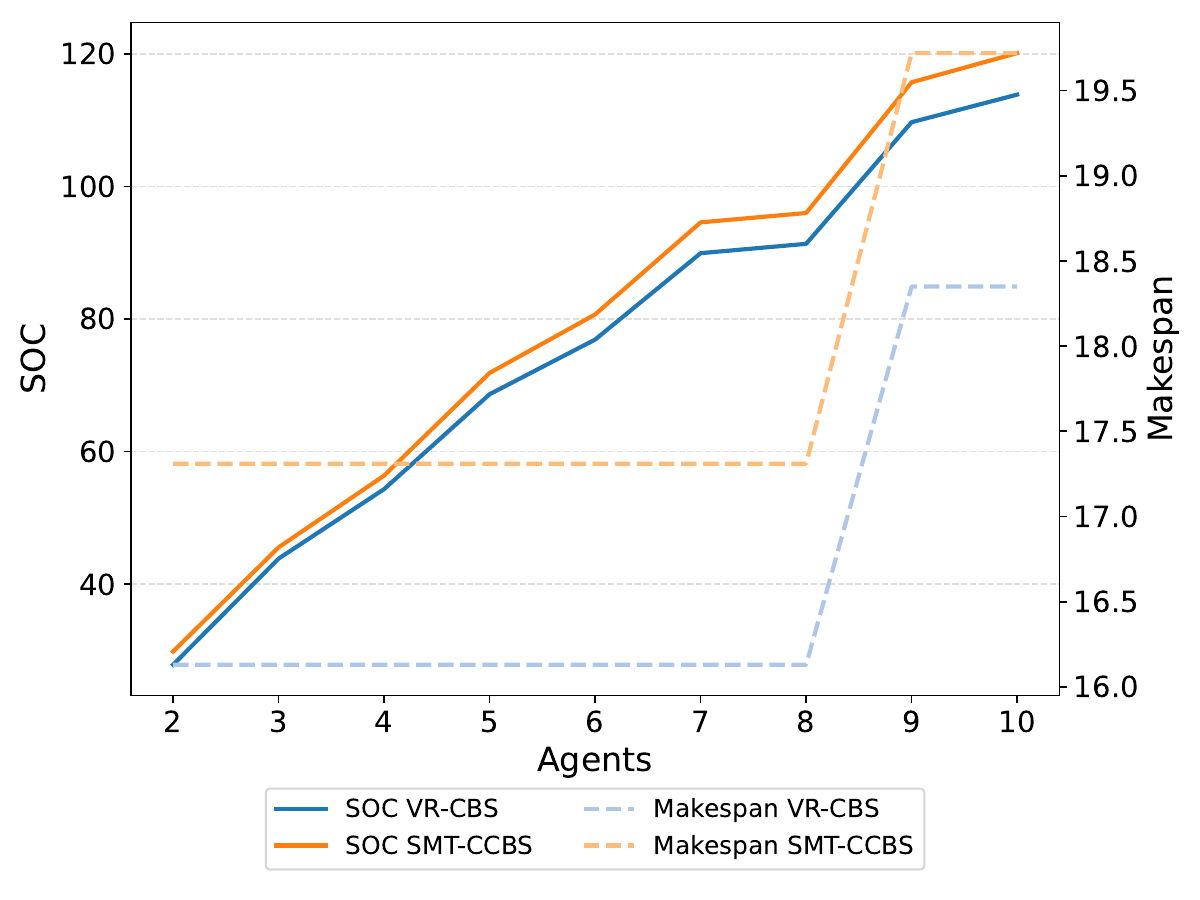} \\[0.8em]
\hline
\multicolumn{2}{|c||}{\textbf{den407d}} & \multicolumn{2}{c|}{\textbf{den408d}} \\[0.3em]
\hline
\includegraphics[width=0.14\textwidth]{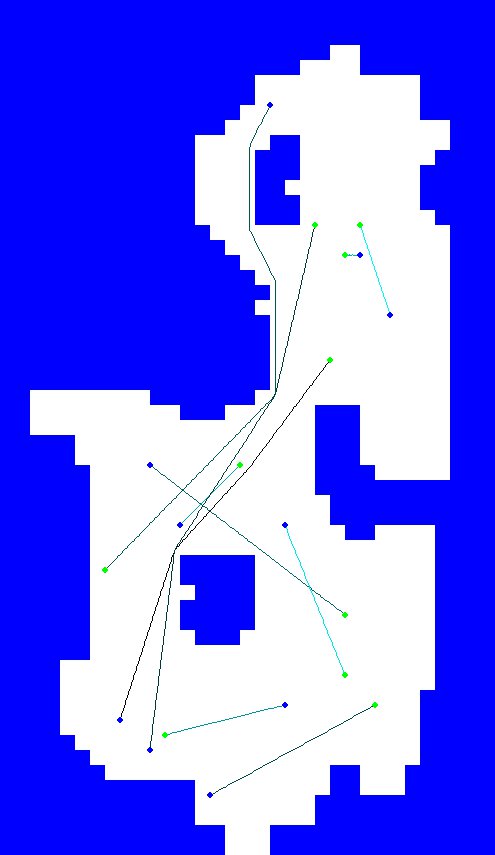} &
\includegraphics[width=0.34\textwidth]{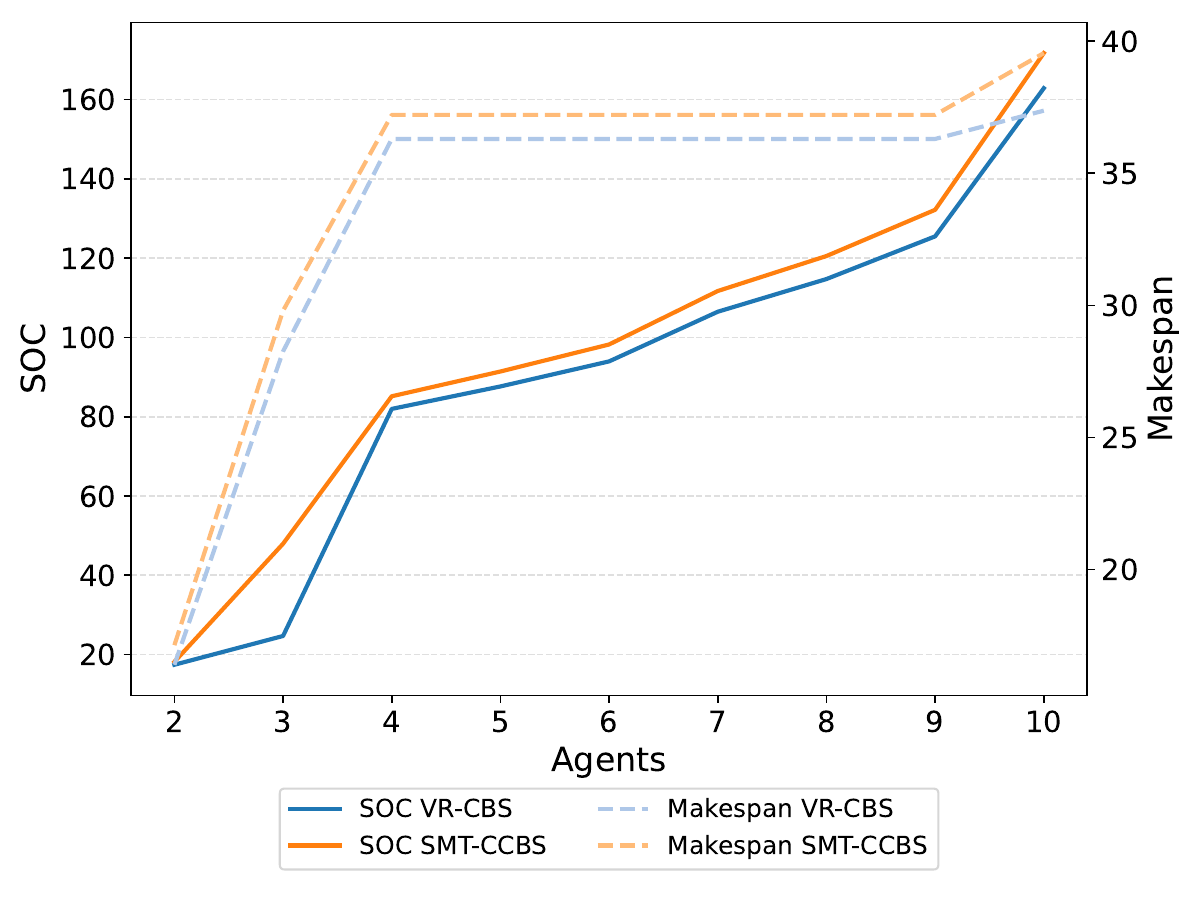} &
\includegraphics[width=0.14\textwidth]{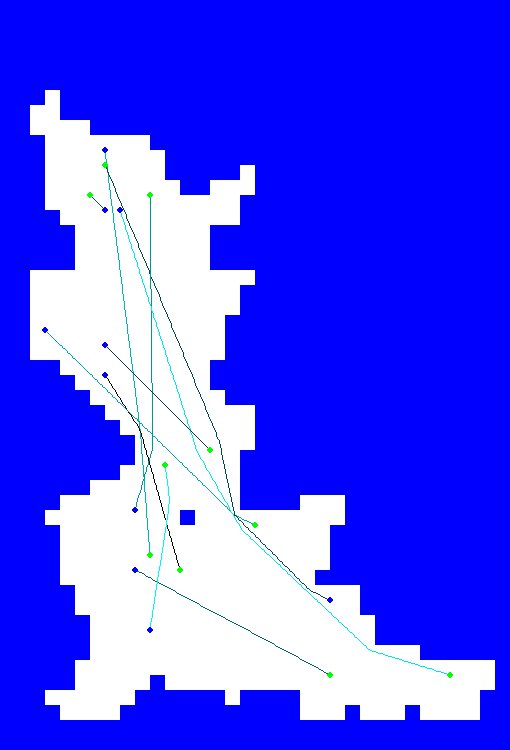} &
\includegraphics[width=0.34\textwidth]{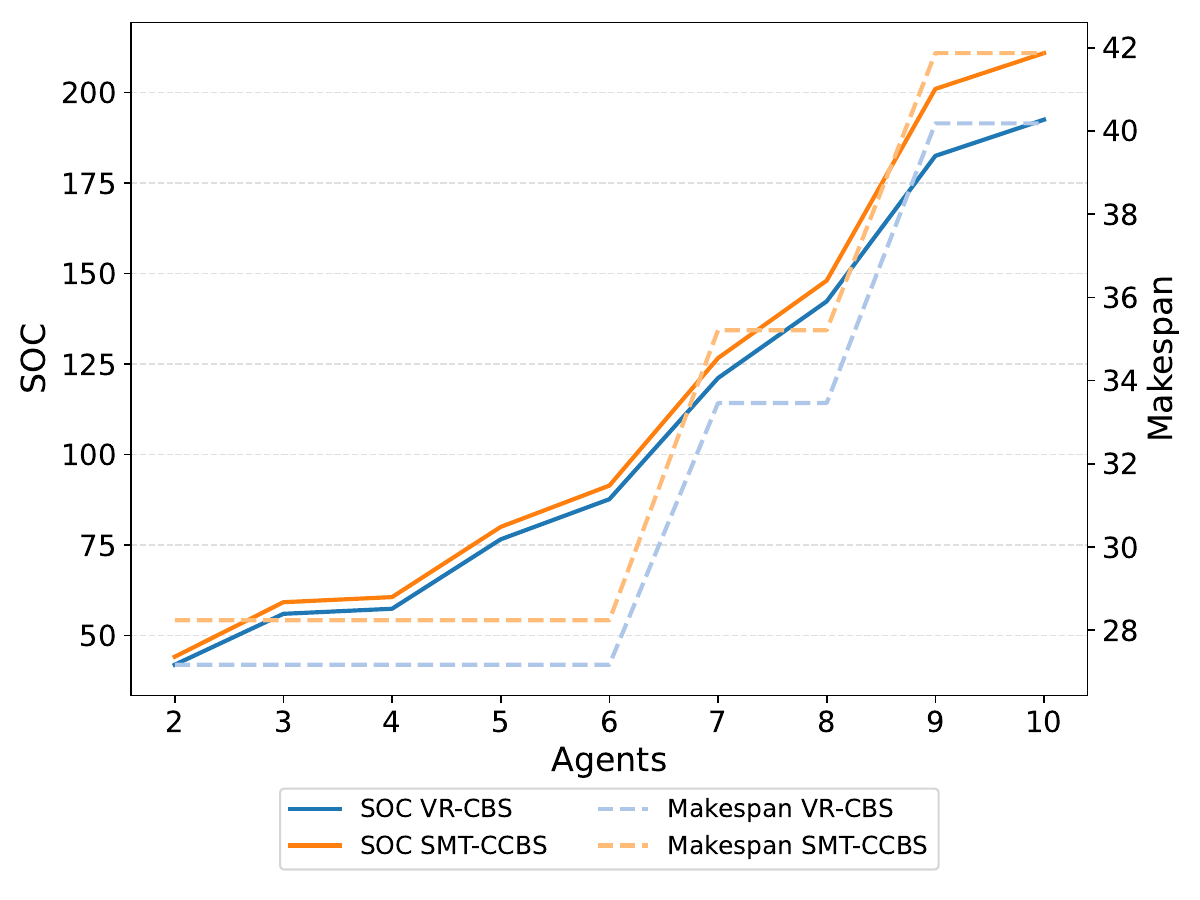} \\[0.8em]
\hline
\multicolumn{2}{|c||}{\textbf{hrt002d}} & \multicolumn{2}{c|}{\textbf{lak101d}} \\[0.3em]
\hline
\includegraphics[width=0.14\textwidth]{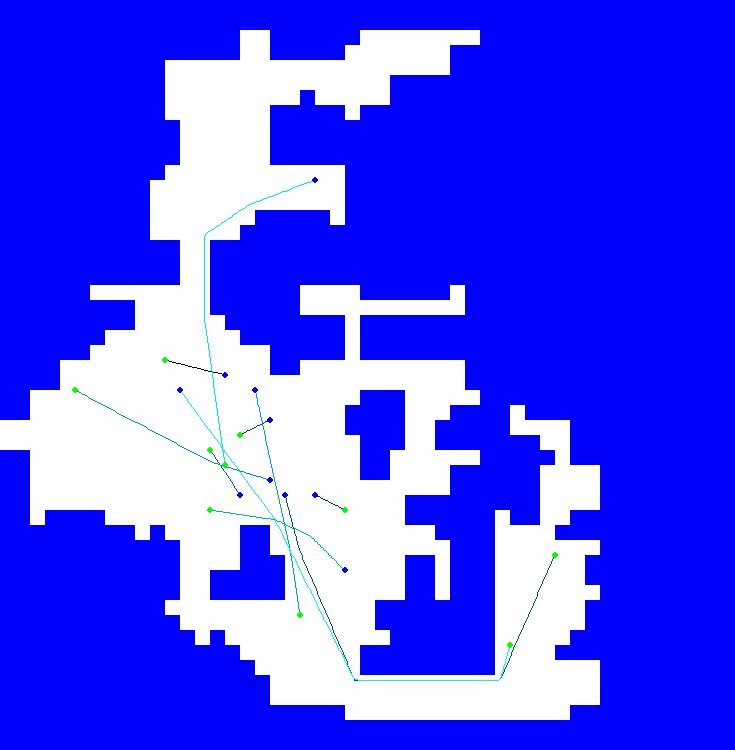} &
\includegraphics[width=0.34\textwidth]{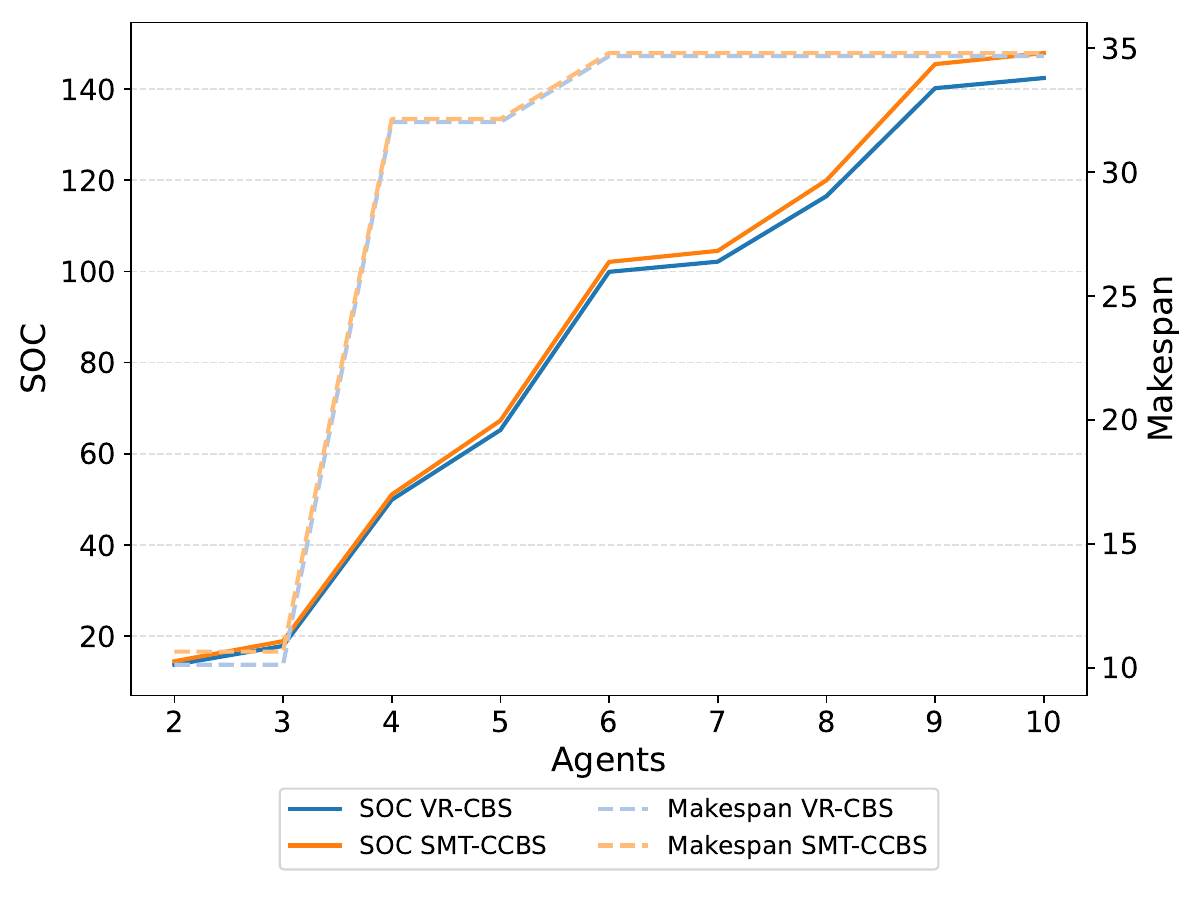} &
\includegraphics[width=0.14\textwidth]{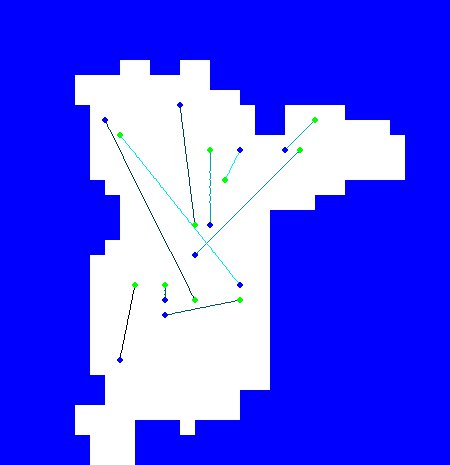} &
\includegraphics[width=0.34\textwidth]{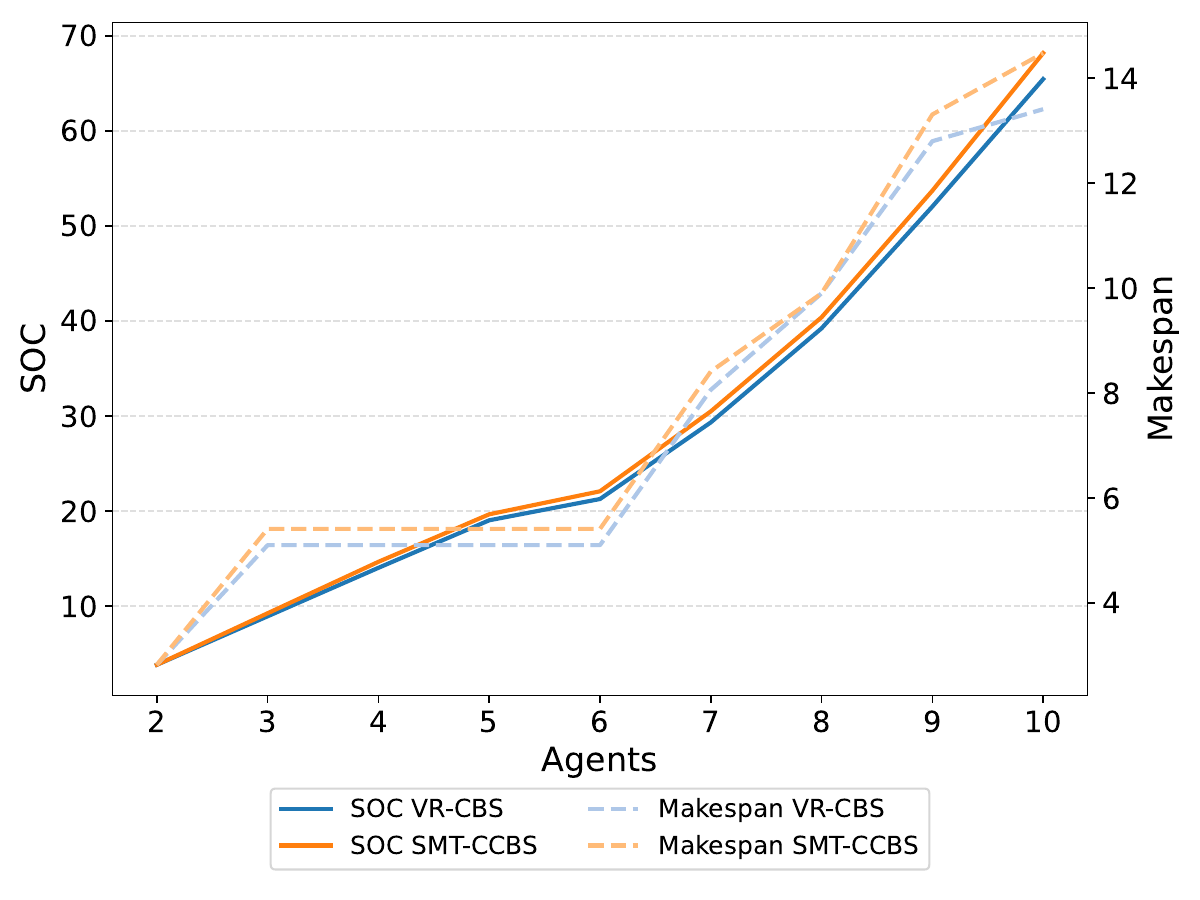} \\[0.8em]
\hline
\multicolumn{2}{|c||}{\textbf{lak102d}} & \multicolumn{2}{c|}{\textbf{lak107d}} \\[0.3em]
\hline
\includegraphics[width=0.14\textwidth]{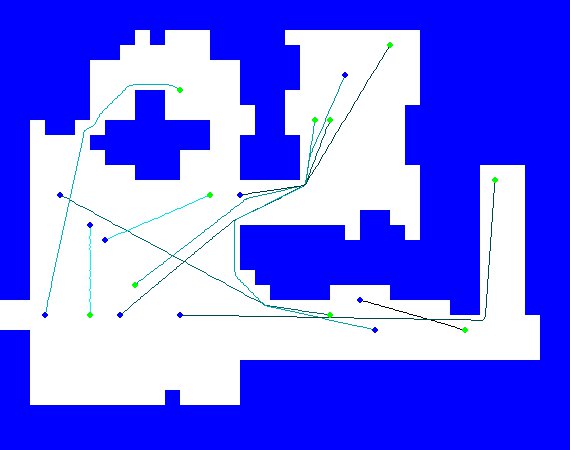} &
\includegraphics[width=0.34\textwidth]{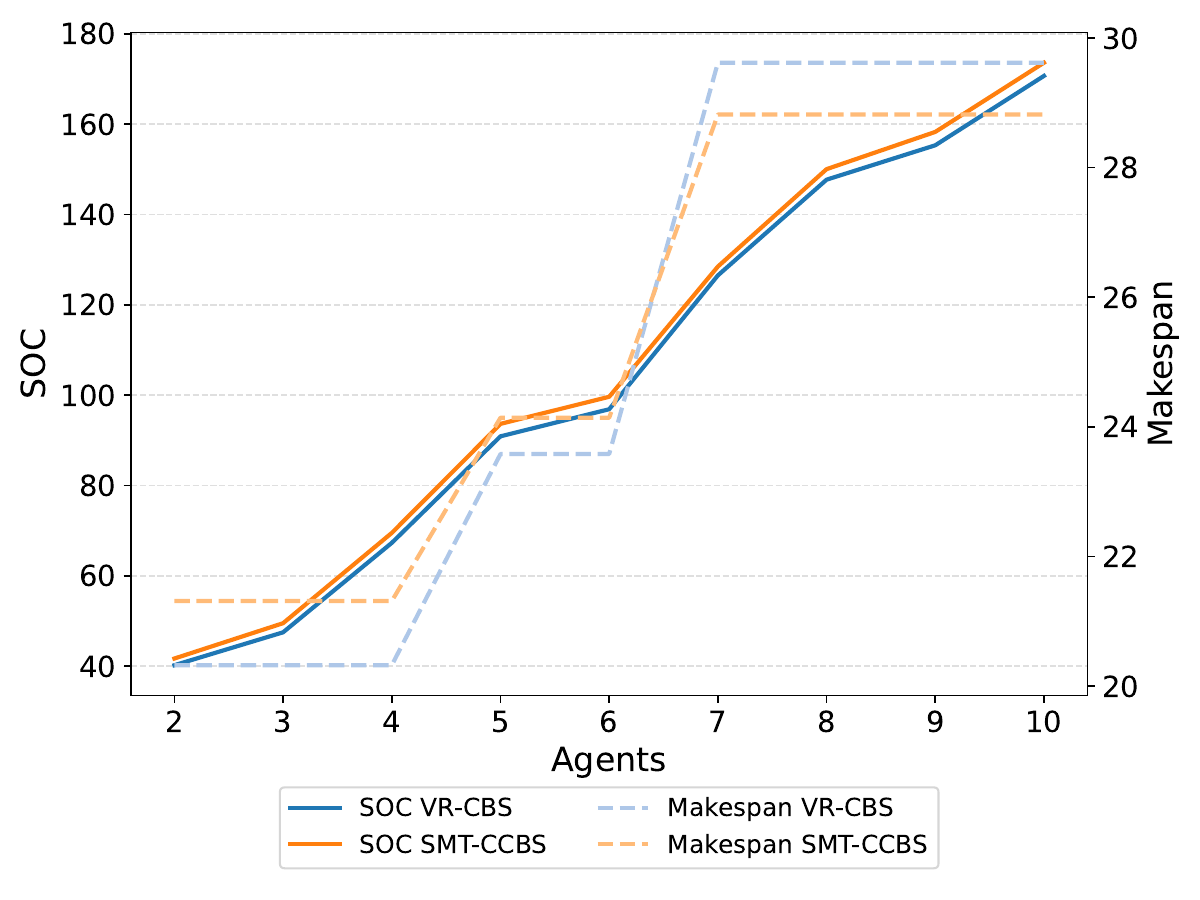} &
\includegraphics[width=0.14\textwidth]{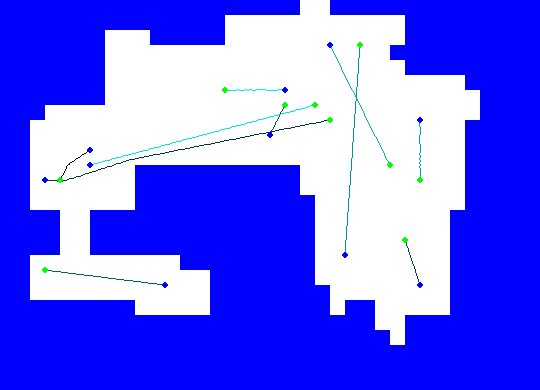} &
\includegraphics[width=0.34\textwidth]{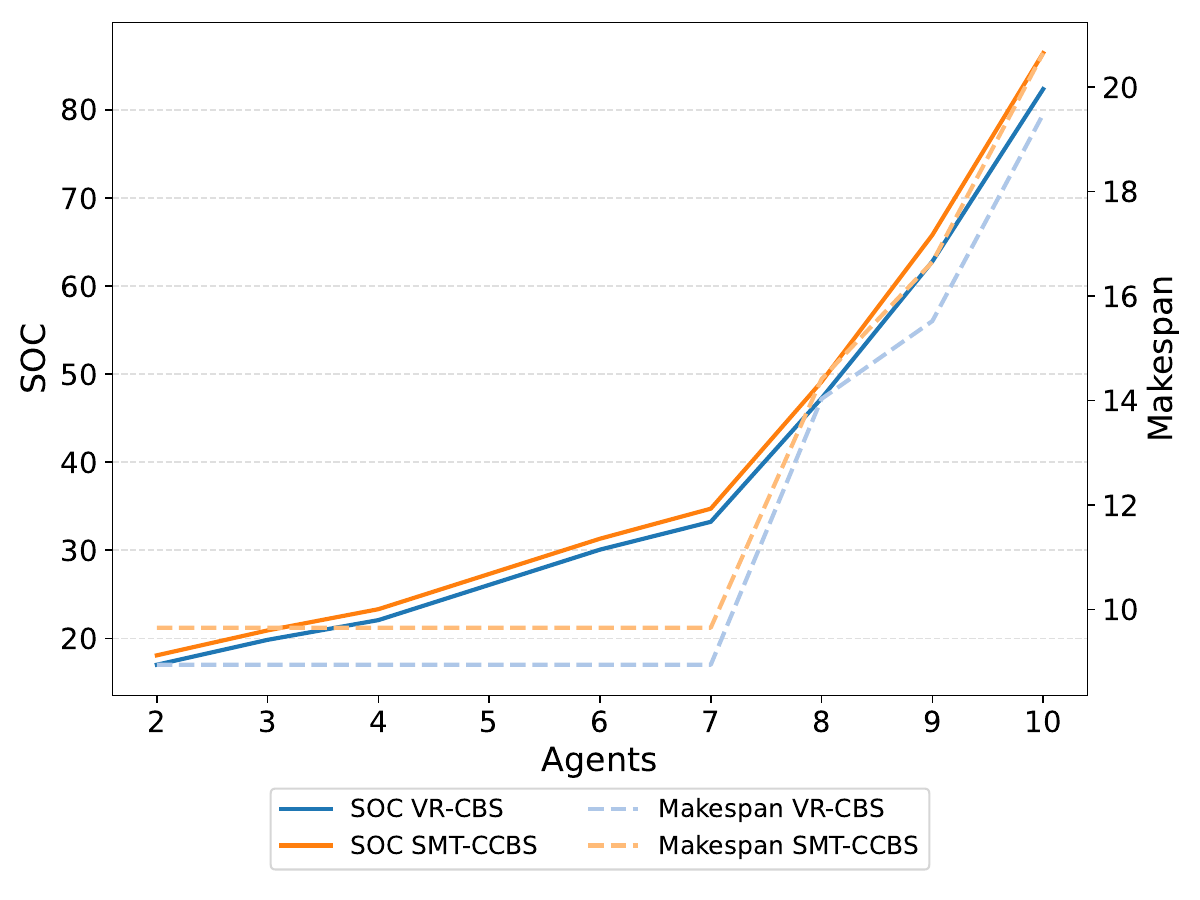} \\[0.8em]
\hline
\end{tabular}
\caption{Comparison of VR-CBS and SMT-CCBS on 8 instances from the DAO grid map data set \cite{movingai}, each a different map with 1 - 10 agents. Agents positions were generated randomly.}
\label{tab:smt1}
\end{table*}

\begin{table*}[p]
\centering
\renewcommand{\arraystretch}{1.2}
\setlength{\tabcolsep}{2pt}
\begin{tabular}{| m{0.14\linewidth} | m{0.34\linewidth} || m{0.14\linewidth} | m{0.34\linewidth}|}
\hline
\multicolumn{2}{|c||}{\textbf{lak108d}} & \multicolumn{2}{c|}{\textbf{lak104d}} \\[0.3em]
\hline
\includegraphics[width=0.14\textwidth]{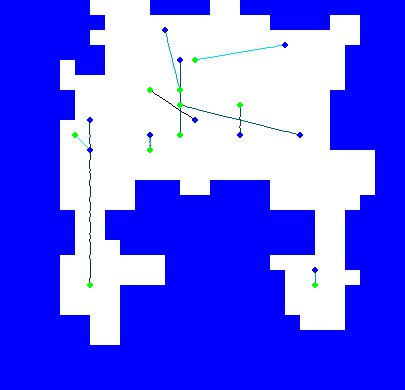} &
\includegraphics[width=0.34\textwidth]{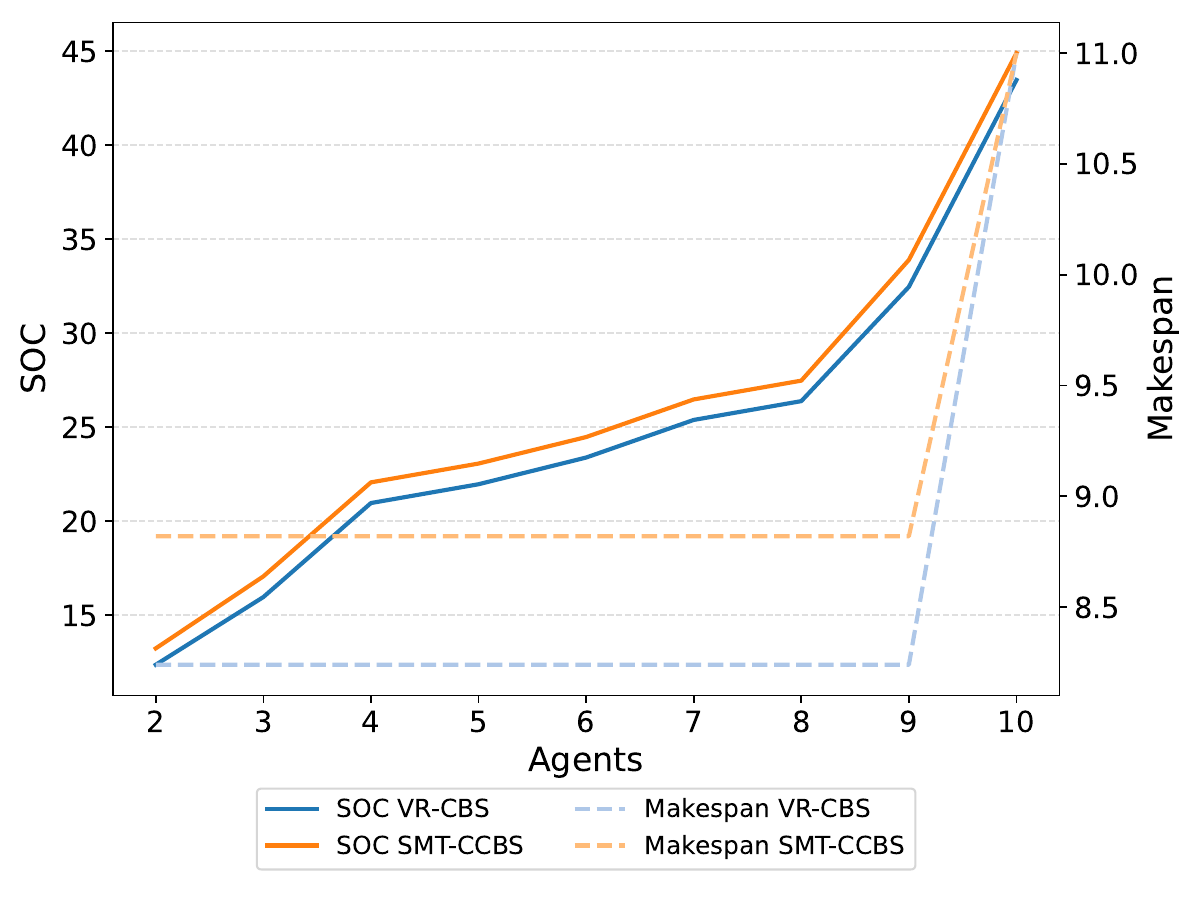} &
\includegraphics[width=0.14\textwidth]{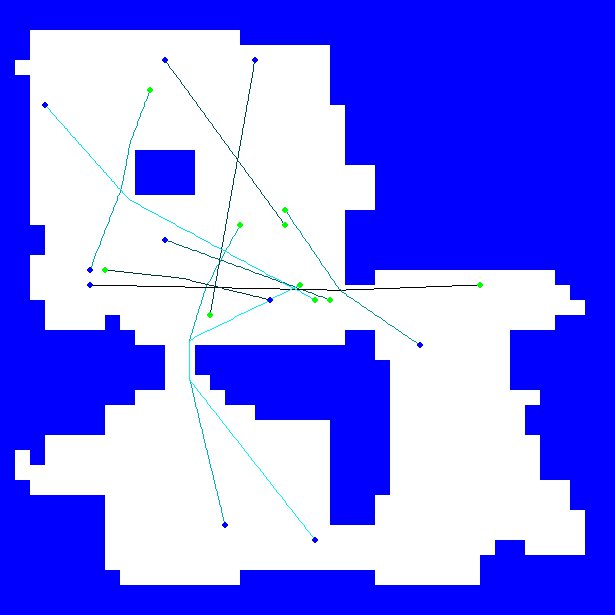} &
\includegraphics[width=0.34\textwidth]{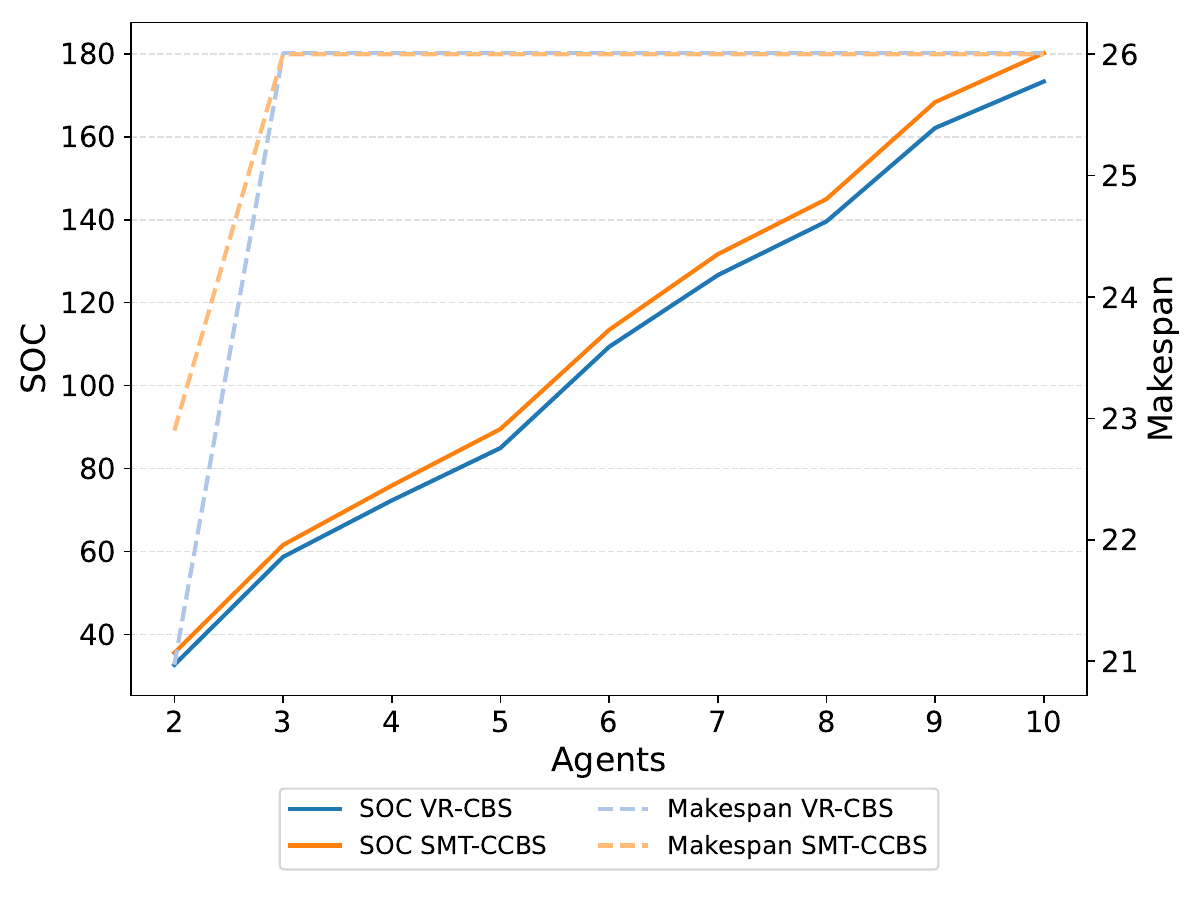} \\[0.8em]
\hline

\multicolumn{2}{|c||}{\textbf{orz102d}} & \multicolumn{2}{c|}{\textbf{orz105d}} \\[0.3em]
\hline
\includegraphics[width=0.14\textwidth]{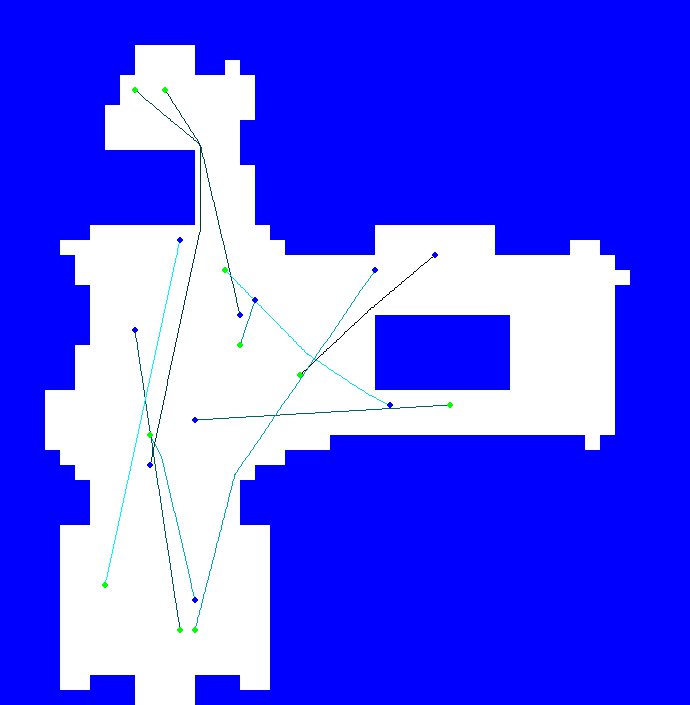} &
\includegraphics[width=0.34\textwidth]{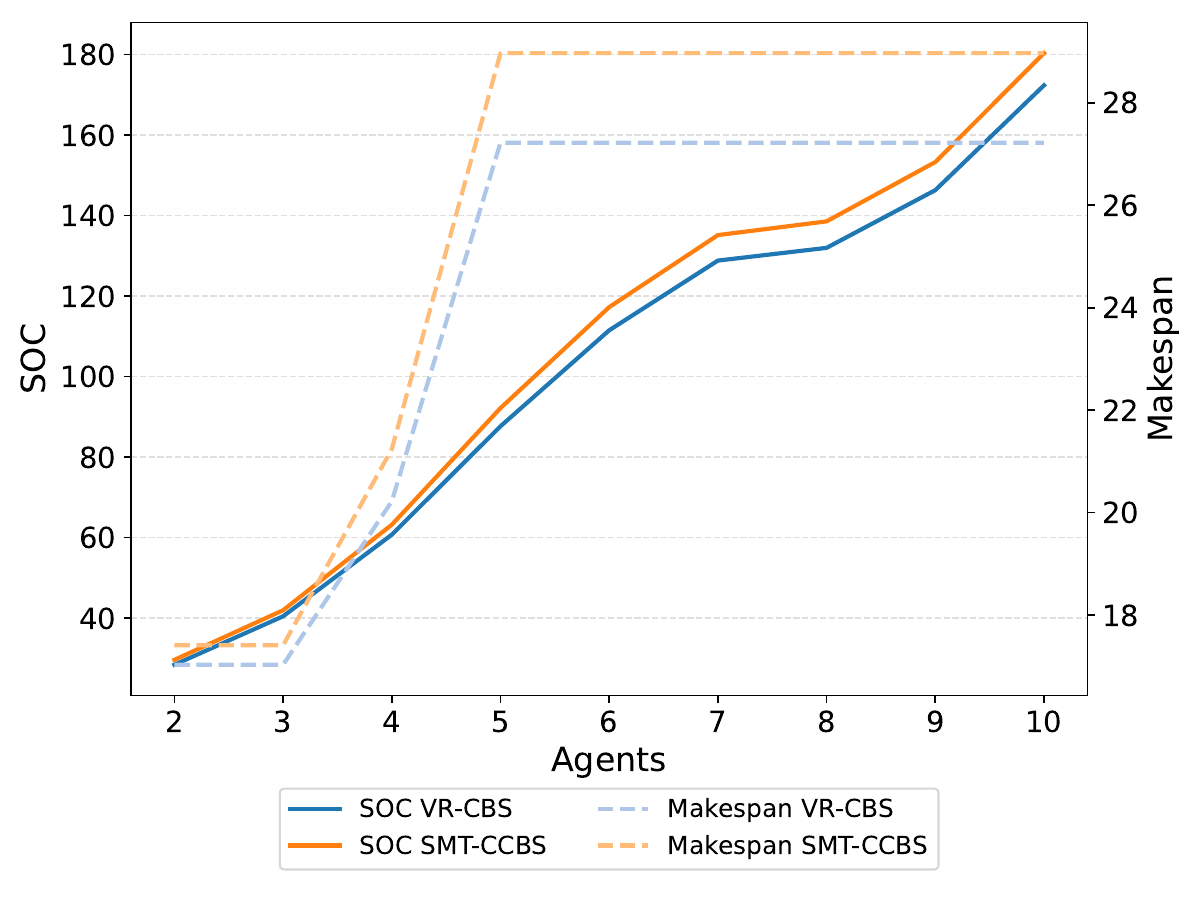} &
\includegraphics[width=0.14\textwidth]{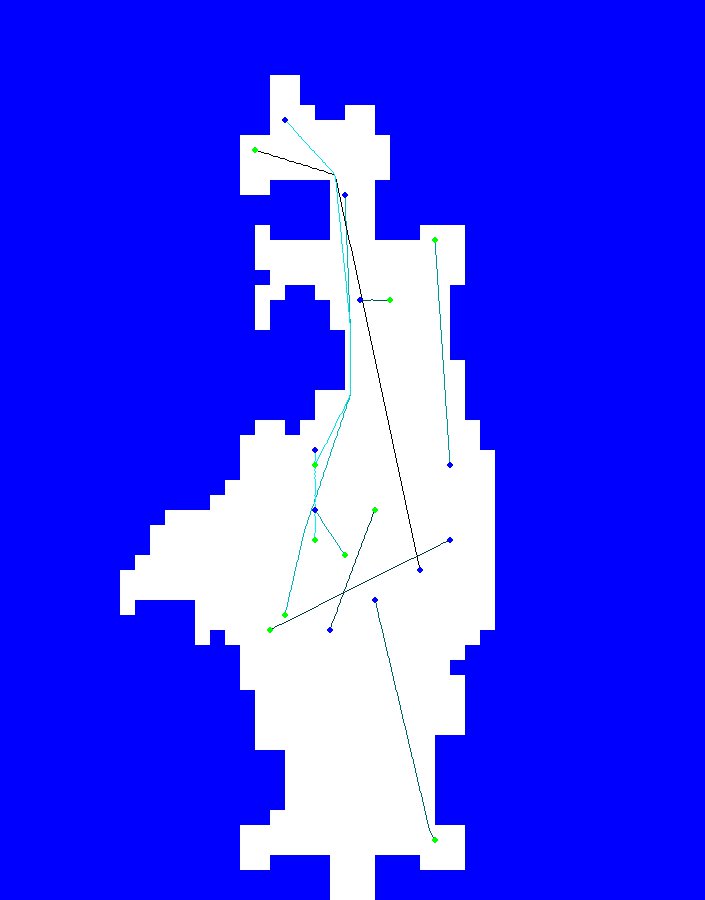} &
\includegraphics[width=0.34\textwidth]{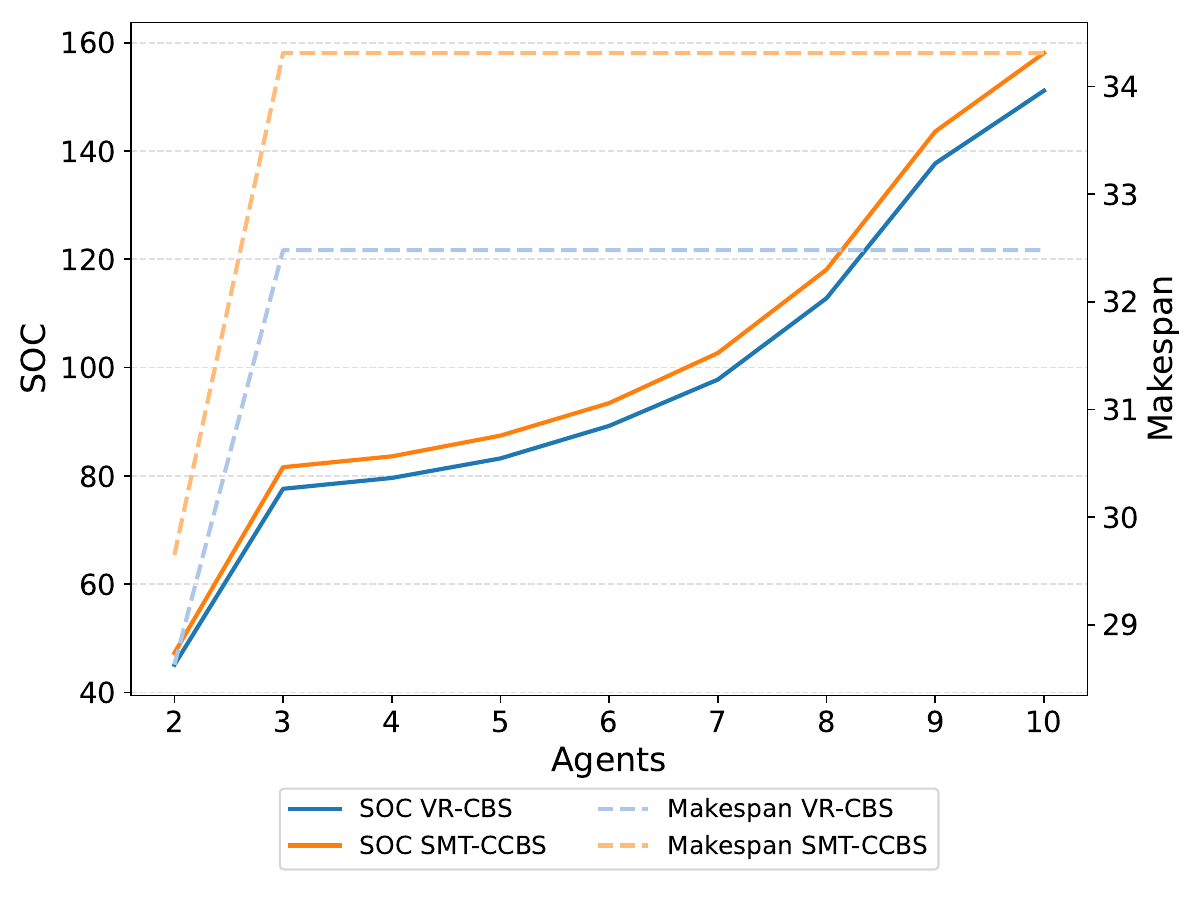} \\[0.8em]
\hline
\multicolumn{2}{|c||}{\textbf{orz107d}} & \multicolumn{2}{c|}{\textbf{orz601d}} \\[0.3em]
\hline
\includegraphics[width=0.14\textwidth]{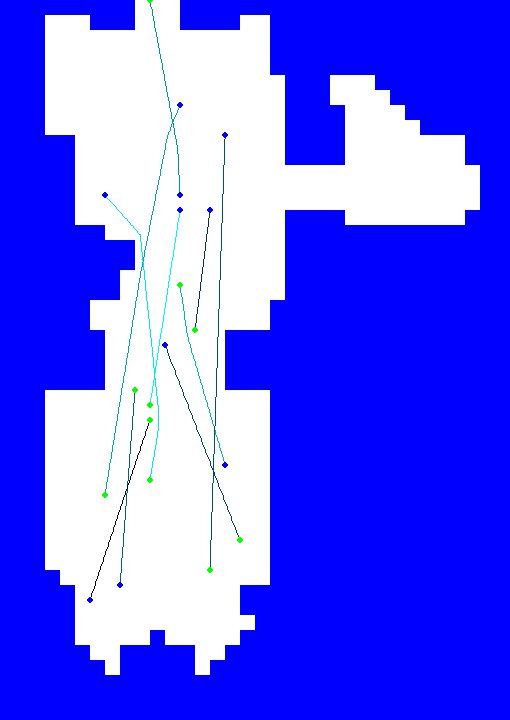} &
\includegraphics[width=0.34\textwidth]{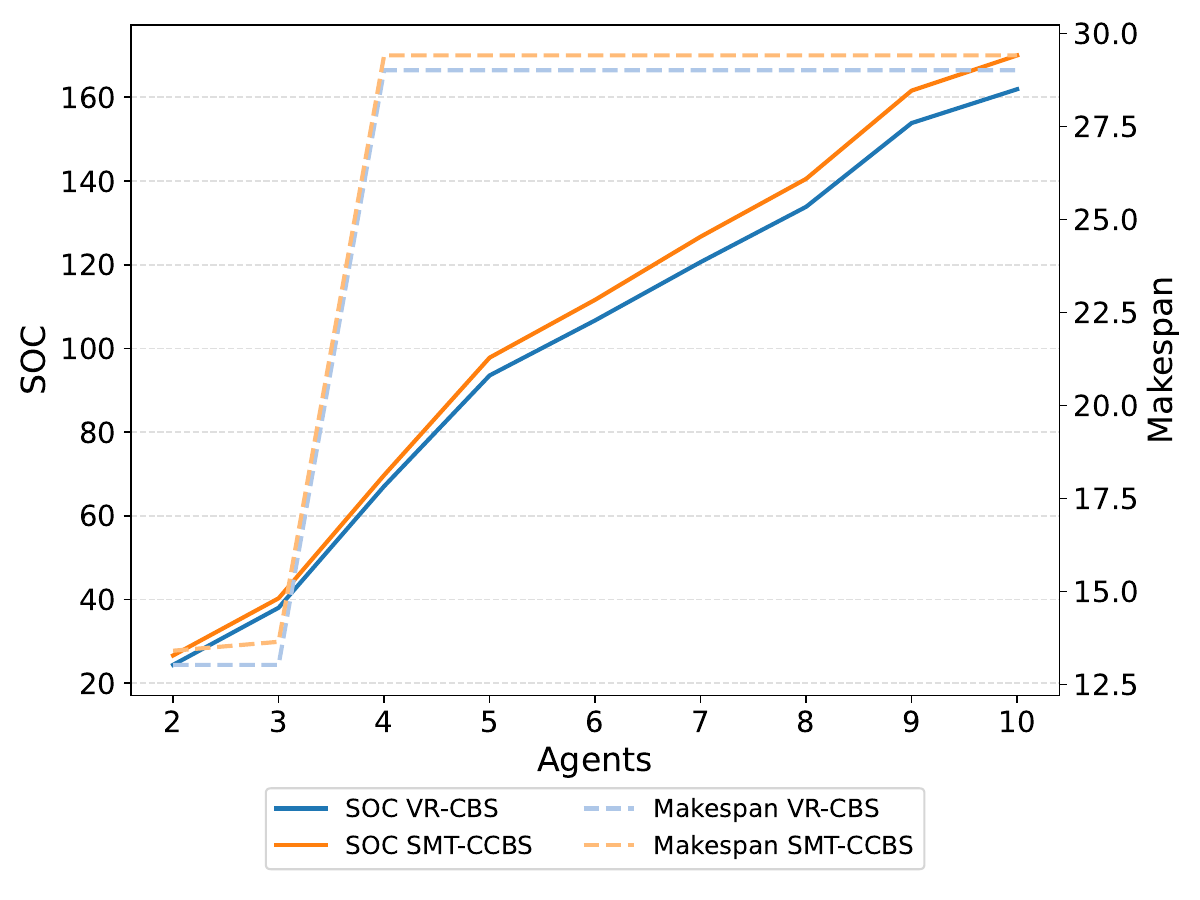} &
\includegraphics[angle=90,width=0.11\textwidth]{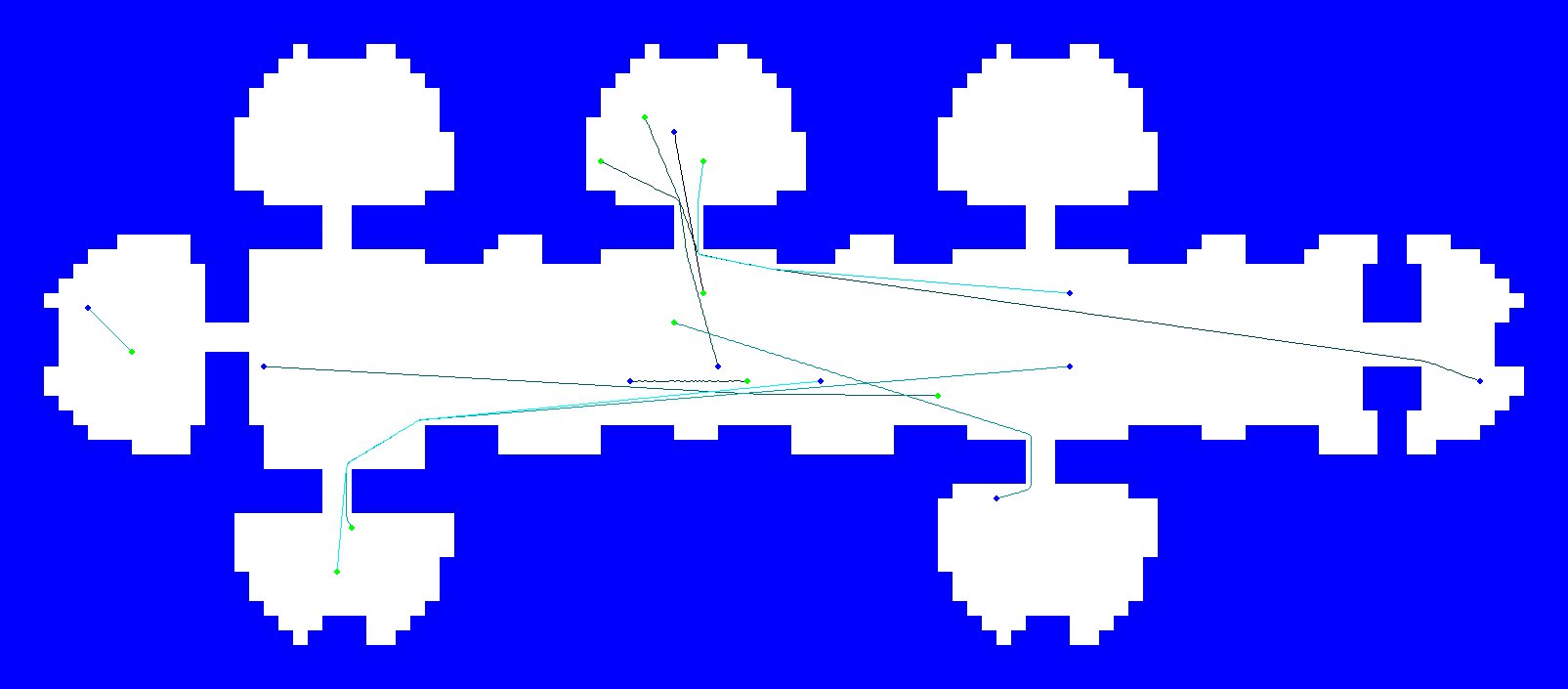} &
\includegraphics[width=0.34\textwidth]{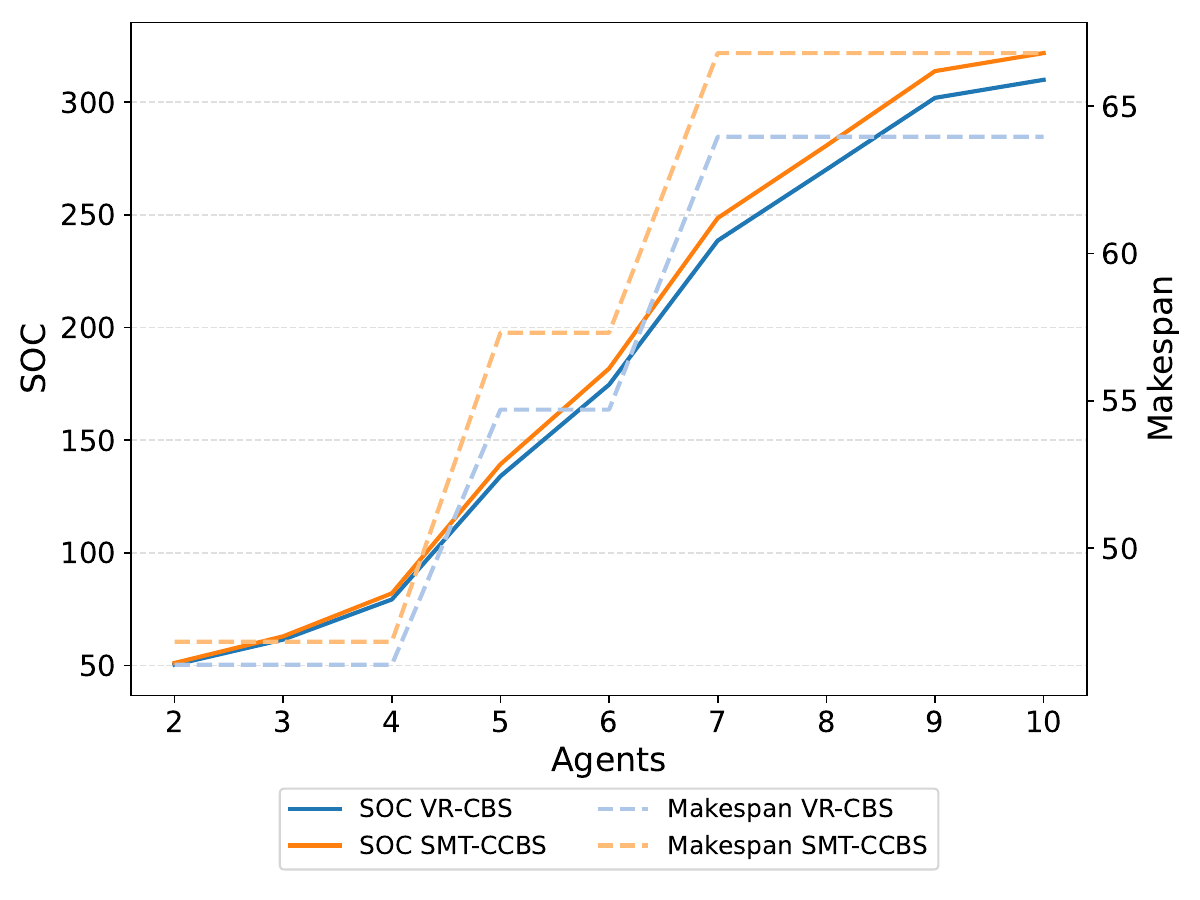} \\[0.8em]
\hline

\multicolumn{2}{|c||}{\textbf{orz704d}} & \multicolumn{2}{c|}{\textbf{lak106d}} \\[0.3em]
\hline
\includegraphics[angle=90,width=0.13\textwidth]{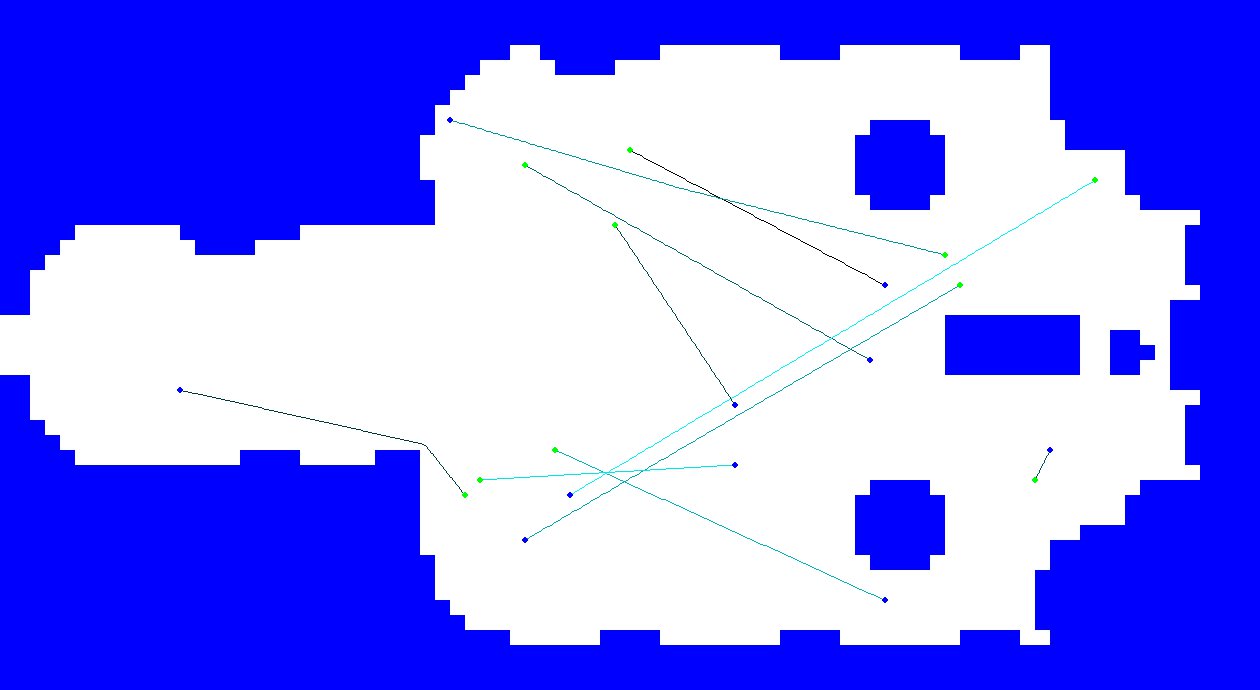} &
\includegraphics[width=0.34\textwidth]{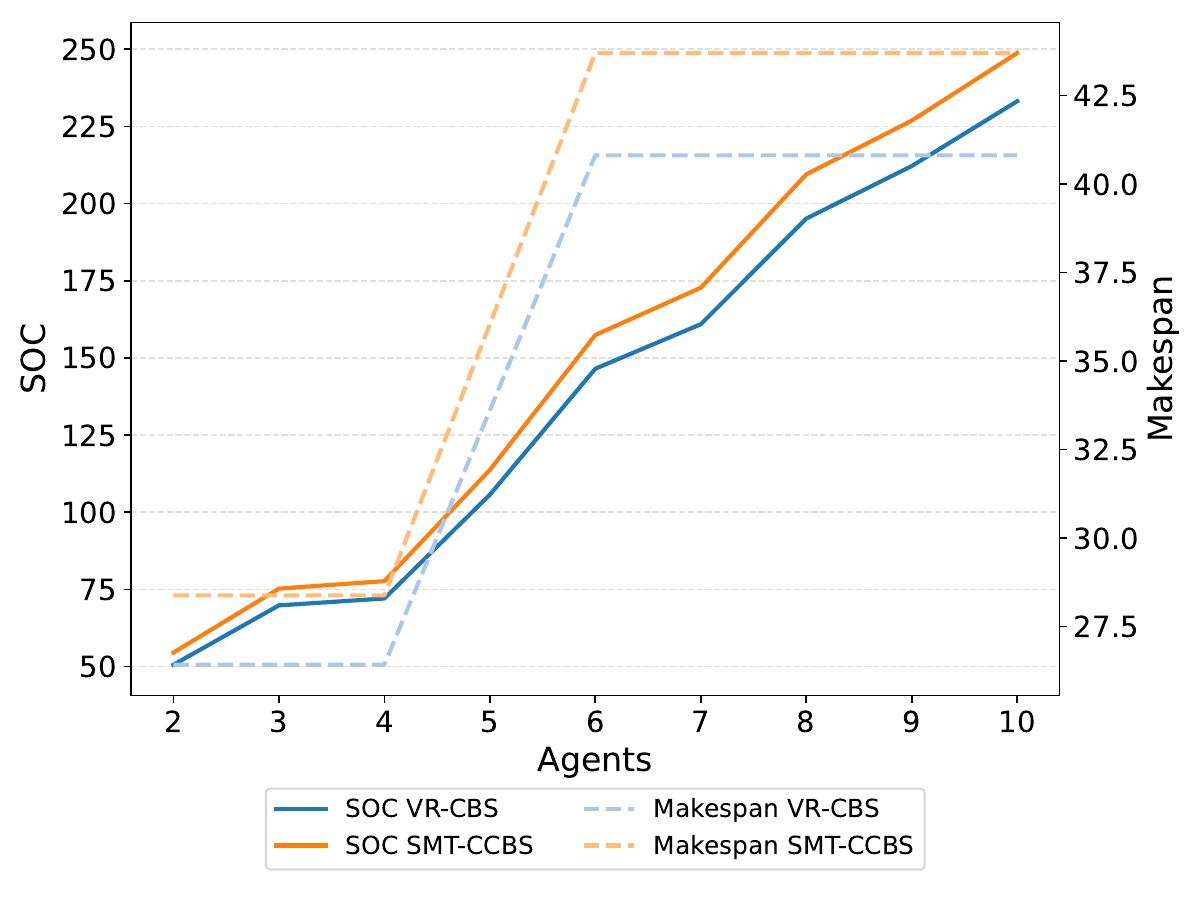} &
\includegraphics[width=0.14\textwidth]{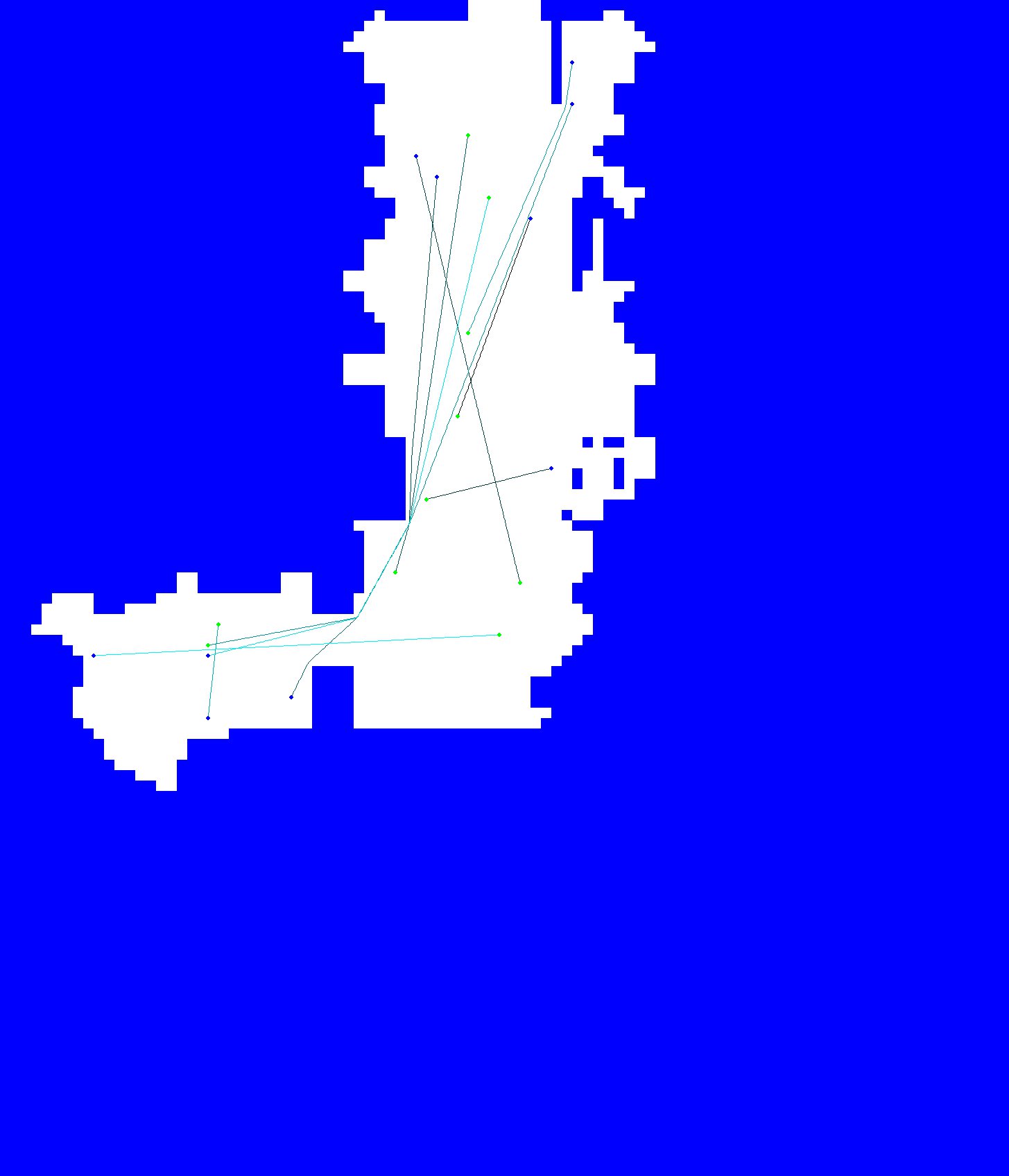} &
\includegraphics[width=0.34\textwidth]{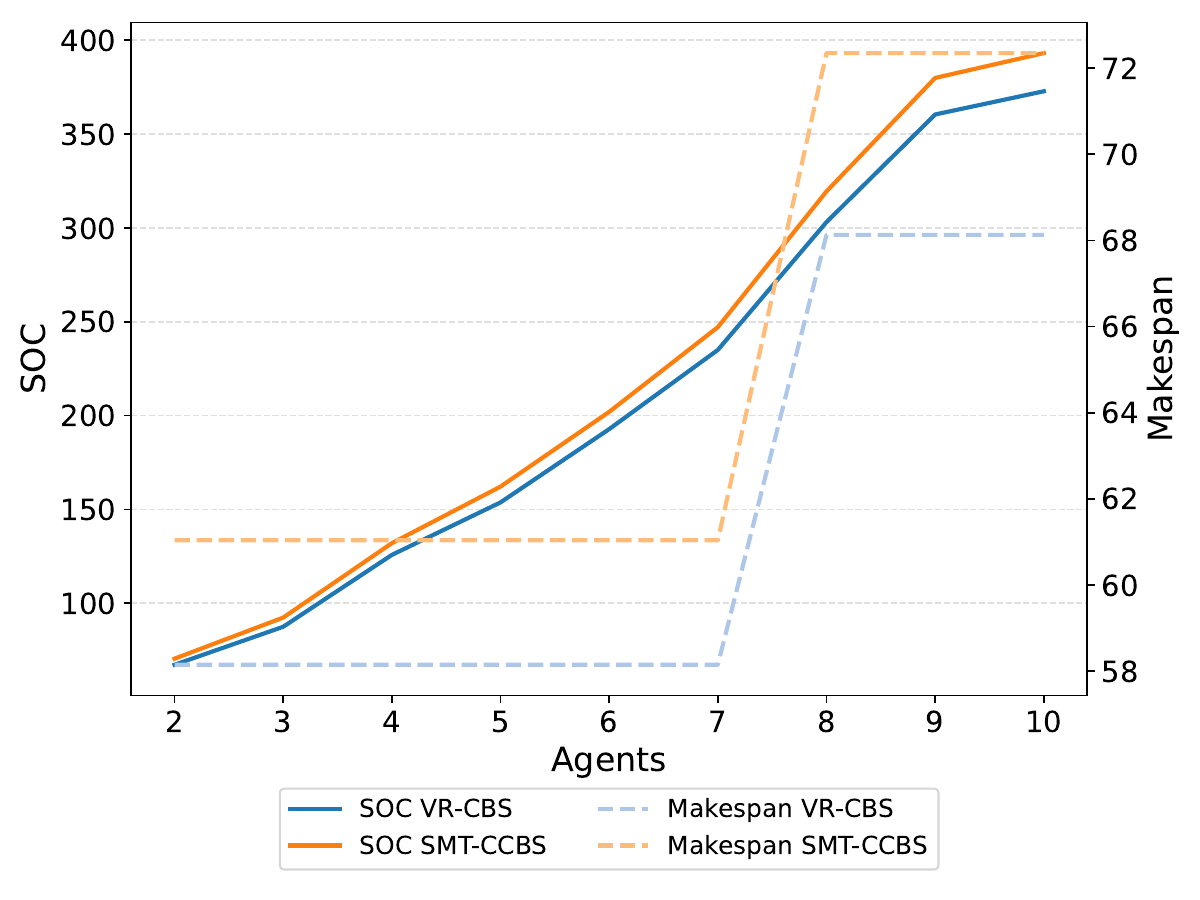} \\[0.8em]
\hline
\end{tabular}
\caption{Comparison of VR-CBS and SMT-CCBS on 8 instances from the DAO grid map data set \cite{movingai}, each a different map with 1 - 10 agents. Agents positions were generated randomly.}
\label{tab:smt2}
\end{table*}

\section{\uppercase{Discussion and conclusion}} This work focusses on multi-agent path finding in a continuous environment. A novel algorithm employing visibility roadmaps, called VR-CBS, is proposed.

VR-CBS combines the CBS-like high-level search of all possible collision eliminations with a new single-agent path finding approach in visibility roadmaps, which are unique for each agent and can change to reflect new collision avoidance constraints. The collision avoidance constraint for the individual agent is translated to a new static obstacle that requires updating the agent's individual visibility roadmap. In addition to this, a path smoothing step is performed to obtain more real robot-friendly solutions. Line segments of which the paths consist are replaced by B-spline curves to eliminate sharp turns.

The experimental evaluation showed that this approach is more efficient in both path quality over CE-CBS and SMT-CCBS and computational time over CE-CBS, as no sampling is done in this model. Experiments have also shown how agents can navigate in many types of environment, including narrow corridors.

In the future, we would like to focus on combinatorial optimization of curve parameters to make them better adjusted to kinematic properties of an agent such as its speed and rate of turn. Another task is to optimize the routes with focus on safer collision avoidance with respect to user requirements.

\section{\uppercase{Acknowledgments}} This work has been supported by the Robotics and Advanced Industrial Production project, registered under number CZ.02.01.01/00/22\_008/0004590.

\bibliographystyle{apalike}
{\small
\bibliography{bibliography}}

\end{document}